\documentclass[journal]{IEEEtran}
\ifCLASSINFOpdf
\else
\fi
\usepackage{amsmath}
\usepackage{amssymb}

\usepackage[ruled,linesnumbered]{algorithm2e}

\usepackage[utf8]{inputenc}
\usepackage{url}
\usepackage{breakurl}
\usepackage{color}
\usepackage{etoolbox}
\usepackage{setspace}

\usepackage[table]{xcolor}
\usepackage{multirow}
\usepackage{hhline}
\usepackage{makecell, cellspace}
\makeatletter
\patchcmd{\@algocf@start}
  {-1.5em}
  {0pt}
  {}{}
\makeatother

\usepackage{pgfplots}
\pgfplotsset{compat=newest}
\usetikzlibrary{patterns}
\newlength\figurewidth
\newlength{\hatchspread}
\newlength{\hatchthickness}
\newlength{\hatchshift}
\newcommand{\hatchcolor}{}
\tikzset{hatchspread/.code={\setlength{\hatchspread}{#1}},
         hatchthickness/.code={\setlength{\hatchthickness}{#1}},
         hatchshift/.code={\setlength{\hatchshift}{#1}},
         hatchcolor/.code={\renewcommand{\hatchcolor}{#1}}}
\tikzset{hatchspread=3pt,
         hatchthickness=0.4pt,
         hatchshift=0pt,
         hatchcolor=black}
\pgfdeclarepatternformonly[\hatchspread,\hatchthickness,\hatchshift,\hatchcolor]
   {custom north west lines}
   {\pgfqpoint{\dimexpr-2\hatchthickness}{\dimexpr-2\hatchthickness}}
   {\pgfqpoint{\dimexpr\hatchspread+2\hatchthickness}{\dimexpr\hatchspread+2\hatchthickness}}
   {\pgfqpoint{\dimexpr\hatchspread}{\dimexpr\hatchspread}}
   {
    \pgfsetlinewidth{\hatchthickness}
    \pgfpathmoveto{\pgfqpoint{0pt}{\dimexpr\hatchspread+\hatchshift}}
    \pgfpathlineto{\pgfqpoint{\dimexpr\hatchspread+0.15pt+\hatchshift}{-0.15pt}}
    \ifdim \hatchshift > 0pt
      \pgfpathmoveto{\pgfqpoint{0pt}{\hatchshift}}
      \pgfpathlineto{\pgfqpoint{\dimexpr0.15pt+\hatchshift}{-0.15pt}}
    \fi
    \pgfsetstrokecolor{\hatchcolor}
    \pgfusepath{stroke}
   }

\pgfdeclarepatternformonly{newdots}{\pgfqpoint{-1pt}{-1pt}}{\pgfqpoint{0.5pt}{0.5pt}}{\pgfqpoint{1pt}{1pt}}
{%
  \pgfpathcircle{\pgfqpoint{0pt}{0pt}}{.25pt}%
  \pgfusepath{fill}%
}%

\pgfdeclarepatternformonly{newdotss}{\pgfqpoint{-1pt}{-1pt}}{\pgfqpoint{0.25pt}{0.25pt}}{\pgfqpoint{0.5pt}{0.5pt}}
{%
  \pgfpathcircle{\pgfqpoint{0pt}{0pt}}{0.15pt}%
  \pgfusepath{fill}%
}%

\newcommand{\todo}[1]{\textcolor{red}{#1}}

\newcommand{\qm}[1]{``#1''}  

\newcommand{\evaluated}[1]{\textbf{#1}}

\def\RR{\mathbb{R}}

\def\CC{\mathbb{C}}

\def\NN{\mathbb{N}}

\def\x{\vect{x}}
\def\u{\vect{u}}
\def\vv{\vect{v}}
\def\y{\vect{y}}
\def\z{\vect{z}}
\def\w{\vect{w}}

\def\p{\vect{p}}

\def\g{\vect{g}}

\def\d{\vect{d}}

\def\Mr{M_\mathrm{R}}
\def\Mrt{M_{\mathrm{R}t}}
\def\Mh{M_\mathrm{H}}
\def\Ml{M_\mathrm{L}}

\def\One{\vect{1}}

\newcommand{\Id}{\mathrm{Id}}

\newcommand{\ana}{A}

\newcommand{\norm}[1]{\|#1\|}
\newcommand{\abs}[1]{\left\vert#1\right\vert}

\newcommand{\vect}[1]{\mathbf{#1}} 

\newcommand{\argmin}{\mathop{\operatorname{arg\,min}}}
\newcommand{\modulo}{\mathop{\operatorname{mod}}}
\newcommand{\prox}{\mathrm{prox}}
\newcommand{\soft}{\mathrm{soft}}
\newcommand{\clip}{\mathrm{clip}}
\newcommand{\proj}{\mathrm{proj}}

\newcommand{\sgn}{\mathrm{sgn}}
\newcommand{\dEU}{\mathrm{d}}  

\newcommand{\dsdr}{\ensuremath{\Delta\mathrm{SDR}}}
\newcommand{\dsdrc}{\ensuremath{\Delta\mathrm{SDR}_\mathrm{c}}}
\newcommand{\sdrc}{\ensuremath{\mathrm{SDR}_\mathrm{c}}}
\newcommand{\sdr}{\ensuremath{\mathrm{SDR}}}
\newcommand{\Rnonlin}{\ensuremath{R_\mathrm{nonlin}}}

\newcommand{\tc}{\ensuremath{\theta_\mathrm{c}}}

\def\Gammar{\Gamma_\mathrm{R}}
\def\Gammah{\Gamma_\mathrm{H}}
\def\Gammal{\Gamma_\mathrm{L}}

\definecolor{matlabCommentGreen}{RGB}{60,118,61}
\newcommand{\comment}[1]{\textcolor{matlabCommentGreen}{{\hspace{1em}\%\,{#1}}}}

\newcommand{\edt}[1]{{#1}}

\begin{document}
%
\title{A survey and an extensive evaluation of popular audio declipping methods}
%
%
%

\author{Pavel Záviška, 
        Pavel Rajmic, 
        Alexey Ozerov, 
				and Lucas Rencker%
\thanks{P.\ Záviška and P.\,Rajmic are with the Signal Processing Laboratory,  Brno University of Technology, Czech Republic. E-mail: rajmic@vutbr.cz.}
\thanks{A.\,Ozerov is with InterDigital, Cesson-S{\'e}vign{\'e}, France. E-mail: alexey.ozerov@interdigital.com.}
\thanks{L.\,Rencker is with the Centre for Vision, Speech and Signal Processing (CVSSP), University of Surrey, UK. E-mail: l.rencker@surrey.ac.uk. }
\thanks{Manuscript received May 9, 2020; revised October 20, 2020; accepted November 20, 2020.}}

%
%

\markboth{IEEE Journal of Selected Topics in Signal Processing}%
{IEEE Journal of Selected Topics in Signal Processing}
%



\maketitle

\begin{abstract}
    Dynamic range limitations in signal processing often lead to clipping, or saturation, in signals.
		\edt{The task of audio declipping is estimating} the original audio signal, given its clipped measurements, and has attracted much interest in recent years.
		Audio declipping algorithms often make assumptions about the underlying signal, such as sparsity or low-rankness, and about the measurement system. In this paper, we provide an extensive review of audio declipping algorithms proposed in the literature.
    For each algorithm, we present assumptions that are made about the audio signal, the modeling domain, and the optimization algorithm.
    Furthermore, we provide an extensive numerical evaluation of popular declipping algorithms, on real audio data. We evaluate each algorithm in terms of the Signal-to-Distortion Ratio, and also using perceptual metrics of sound quality.
    The article is accompanied by a repository containing the evaluated methods.
\end{abstract}


%
\IEEEpeerreviewmaketitle

\section{Introduction}
%
%
%
%
\IEEEPARstart{C}{lipping} 
is a non-linear signal distortion usually appearing when a~signal exceeds its allowed dynamic range.
As a~typical instance, an analog signal that is digitized can be clipped in value when its original peak values go beyond the largest (or lowest) digit representation. 
For this reason, the effect is also called saturation.

Clipping in audio signals has a great negative effect on the perceptual quality of audio \cite{Tan2003}, and
it reduces the accuracy of automatic speech recognition \cite{Malek2013:Blind.compensation,Tachioka2014:Speech.recog.performance} and other audio analysis applications.
%
To improve the perceived quality of audio, a~recovery of clipped samples can be made;
this process is usually termed \emph{declipping}.

Many audio declipping methods are available today.
They are based on different modeling assumptions,
tested on very different audio datasets, and evaluated by different methodologies.
The goal of the article is to survey the existing approaches to audio declipping,
categorize them and emphasize some interconnections.
A~no less important goal of this contribution is the numerical evaluation of selected audio declipping methods on a~representative dataset and the provision of a~freely available MATLAB toolbox.
The subset of methods under consideration was selected based on
our intention to cover different reconstruction techniques,
on the popularity of the methods, and on the availability---or reproducibility---of their implementation (which go hand in hand in many cases).
It is worth saying that the reconstruction quality is the primary focus in the evaluation,
but comments on the speed of computation are provided as well.

\begin{figure}
    \input{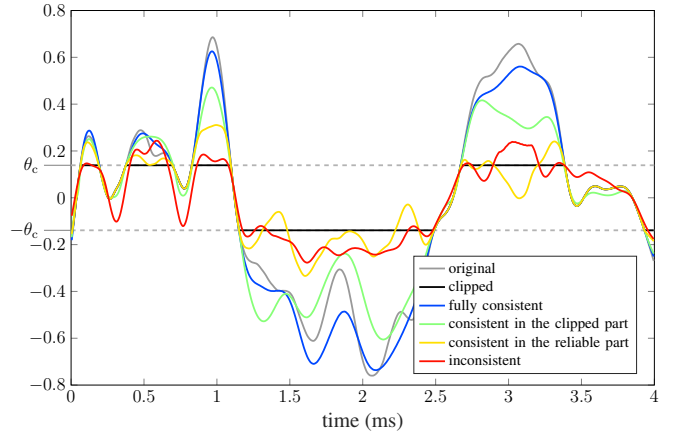}
    \caption{Example of a clipped signal.
				Various recovery possibilities are depicted, showing different types of solution consistency 
        (discussed in Sec.\,\ref{sec:Declipping.methods}).
        }
    \label{fig:waveforms.consistency}
\end{figure}

In the case of the hard clipping, which is the degradation considered in this survey,
the input signal
exceeding the prescribed dynamic range $[-\tc, \tc]$ is limited in amplitude such that
\begin{equation}
y_n = \left\{
\begin{aligned}
&x_n &\text{for} \hspace{1em} &|x_n| < \tc, \\
&\tc \cdot \sgn(x_n) &\text{for} \hspace{1em}  &|x_n| \geq \tc,
\end{aligned}
\right.
\label{eq:clipping}
\end{equation}
where $[x_1,\ldots,x_N] = \x \in \RR^N$ denotes the original (clean) signal and $[y_1,\ldots,y_N] = \y \in \RR^N$ the observed clipped signal.
The limiting constant $\tc$ is referred to as the clipping threshold
(this article supposes that clipping is symmetric, without affecting generality).
See Fig.\,\ref{fig:waveforms.consistency} for an example of a~clipped signal (and its various reconstructions).

Drawing on
the terminology used in audio source separation \cite{ozerov2010multichannel} and machine learning in general \cite{sathya2013comparison},
the audio declipping methods can be cast in two main categories:
\begin{itemize}
    \item
    \emph{unsupervised}, or \emph{blind}, where the signal is recovered assuming some generic regularization (or modeling assumption) on what a~natural audio signal should be like,
    but with no additional
    clean audio signals being involved, 
    \item
    \emph{supervised}, where signal recovery model parameters, or a~part of them, are trained on (estimated from) clean audio examples, which should be similar to the audio sample to be recovered
    (e.g., all signals are speech signals). 
\end{itemize}
To 
date, the vast majority of state-of-the-art audio declipping approaches are unsupervised, and thus, we limit this study to those approaches.
However, supervised approaches are emerging as well, and they are mostly deep neural networks (DNNs)-based, trained to declip speech signals
\cite{Bie2015:Detection.and.Reconstruction.of.speech.for.speaker.recongnition, kashani2019:Image2Image.CNN.Speech.Declipping,MackHabets2019:Declipping.Speech.Using.Deep.Filtering}.
Supervised approaches are more specialized
(since they are usually trained on particular classes of audio),
and \emph{potentially}\/ more powerful,
simply because more relevant information could be contained in the training set.
As such, we believe that supervised learning is one of the potential and promising directions of evolution of research on audio declipping.

As for the unsupervised approaches, they often follow a~generic path:
\begin{enumerate}
    \item A {\it modeling domain} (e.g., time, analysis or synthesis, see Sec.\,\ref{sec:Preliminaries}) is chosen.
    
    \item A generic {\it model} regularizing an audio signal is specified in
    the
    chosen domain (e.g., autoregressive
    model, sparsity, group sparsity or low-rank structure).
    
    \item {\it Model parameters}\/ to be estimated from the clipped signal
    are specified (e.g., decomposition coefficients, coefficients and the dictionary, non-negative matrix factorization
    parameters, etc.).
    
    \item A {\it criterion}\/ linking the model parameters and observations to be optimized is specified
    (though the criterion is related to modeling, different choices are often possible; for instance, sparsity-assisted methods may penalize coefficients with the $\ell_0$ or $\ell_1$ norm).
    The criterion may or may not
    include the following ingredients:
    \begin{itemize}
        \item {\it Clipped part consistency}: whether the clipping constraint in the missing part holds
				(see Sec.\,\ref{sec:Declipping.methods}).
        \item {\it Reliable part consistency}: whether the reconstructed signal in the reliable
        part equals the observed clipped signal (see Sec.\,\ref{sec:Declipping.methods}).
    \end{itemize}
    
    \item A suitable {\it algorithm}
    to optimize the model criterion is chosen\,/\,constructed
    (e.g., orthogonal matching pursuit,
    expectation maximization,
    etc.).
    
    \item Once the algorithm has terminated (typically, a~fixed number of iterations are performed or a condition designed to check convergence is satisfied),
    the final signal is formed.
\end{enumerate}
Most of state-of-the-art unsupervised audio declipping approaches characterized by the above-mentioned ingredients,
including the approaches evaluated in this paper,
are summarized in Table~\ref{tab::methods_summary}.

In the case of multichannel audio (e.g., stereo) declipping may exploit correlations between different audio sources\,/\,objects in different channels,
and this can improve the result over a~straightforward, dummy solution of applying a~single-channel declipping to each channel independently.
This was for the first time investigated in \cite{OzerovBilenPerez2016:MultichannelAudioDeclipping},
and then studied in \cite{GaultierBertinGribonval2018:CASCADE} as well,
though with a~different approach.
These works have shown
that using the inter-channel dependencies
can indeed improve the declipping performance.
We do not evaluate those methods in this article, though,
since there are only a few of them so far and such a~task would require the creation of a~particular multichannel dataset.

\begin{table*}[t]
\begin{center}
\footnotesize
\caption{%
Categorization of existing single-channel unsupervised declipping approaches.
Methods treated in detail and evaluated are highlighted in bold.
%
Table entries that did not fit into a~clear category are left with
the \qm{N/A} mark.
}
\begin{tabular}{|Sl||Sc|Sc|Sc|Sc|Sc|Sc|Sc|}
  \hline
	\rowcolor{white!90!black}
  Method & \makecell{Modeling\\domain} &  \makecell{Modeling\\assumptions}  &   \makecell{Model\\parameters}   & \makecell{Optimization\\criterion} &  \makecell{Clipping\\consistency}   &  \makecell{Rel.\ part\\consistency}  & \makecell{Optimization\\algorithm} \\

  \hhline{|=::=======|}
  \evaluated{Janssen'86} \cite{javevr86} & AR & AR model & AR params. & ML & no & yes & EM \\
	\rowcolor{white!95!black}
  Abel'91 \cite{abel91_declipping} & spectrum & \makecell{limited\\bandwidth} & band limit & several & yes & yes & N/A \\
  Fong'01 \cite{FongGodsill2001:MonteCarlo} & N/A & AR model & N/A & \makecell{AR coefs \&\\correlation coefs} & N/A & N/A & Monte Carlo \\
  %
  %
	\rowcolor{white!95!black}
  Dahimene'08 \cite{Dahimene2008_declipping} & time & AR model & AR params. & least squares & no & yes & pseudoinverse \\
  \evaluated{Adler'11} \cite{Adler2011:Declipping} & synthesis & sparsity & transform coefs & $\ell_0$-min & yes & no & OMP \\ 
	\rowcolor{white!95!black}
  \evaluated{Weinstein'11}\,\cite{Weinstein2011:DeclippingSparseland} & sparsity & sparsity & transform coefs & reweighted $\ell_1$-min & yes & yes & CVX \\ 
  Miura'11 \cite{Miura2011:DeclippingRecursiveVectorProjection} & synthesis & sparsity & transform coefs & $\ell_0$-min & no & N/A & RVP (MP) \\ 
	\rowcolor{white!95!black}
  Kiti\'c'13 \cite{Kitic2013:Consistent.iter.hard.thresholding} & synthesis & sparsity & transform coefs & $\ell_0$-min & approximate & approximate & IHT \\
  \evaluated{Defraene'13} \cite{Defraene2013:Declipping.perceptual.compressed.sensing} & synthesis & \makecell{sparsity \&\\ psychoacoust.} & transform coefs & $\ell_1$-min & yes & no & CVX \\
  %
	%
	\rowcolor{white!95!black}
	Selesnick'13 \cite{Selesnick2013:LeastSquares} & time & smoothness & signal samples & regularized LS & no & no & explicit formula \\
  \evaluated{Siedenburg'14}\,\cite{SiedenburgKowalskiDoerfler2014:Audio.declip.social.sparsity} & synthesis & social sparsity & transform coefs &  social shrinkage & approximate & approximate & (F)ISTA \\
	\rowcolor{white!95!black}
  Kiti\'c'14 \cite{KiticBertinGribonval2014:AudioDeclippingCosparseHardThresholding} & analysis & sparsity & transform coefs &  $\ell_0$-min & yes & yes & ADMM \\ 
  Jonscher'14 \cite{Jonscher2014:DeclippingFSE} & synthesis & sparsity & transform coefs & N/A & no & N/A & N/A \\ 
	\rowcolor{white!95!black}
  \evaluated{Bilen'15} \cite{BilenOzerovPerez2015:declipping.via.NMF} & analysis & low-rank NMF & NMF params. & ML & yes & yes & EM \\ 
  \evaluated{Kiti\'c'15} \cite{Kitic2015:Sparsity.cosparsity.declipping} & \makecell{analysis \&\\ synthesis} & sparsity & transform coefs & $\ell_0$-min & yes & yes & ADMM \\
	\rowcolor{white!95!black}
  Harvilla'15 \cite{HarvillaStern2015:RBAR} & time & smoothness & signal samples & regularized LS & no & no & explicit formula \\
  Takahashi'15 \cite{Takahashi2015:Block.Adaptive.Decipping.Null.Space} & N/A & low rank & signal samples & quadratic & yes & yes & custom \\ 
	\rowcolor{white!95!black}
  Elvander'17 \cite{Elvander2017:Gridless.estimation.saturated.signals} & synthesis & sparsity & transform coefs & atomic norm min & yes & yes & SD \\ 
  %
    %
  \evaluated{Rencker'18} \cite{RenckerBachWangPlumbley2018:Consistent.dictionary.learning.LVA} & synthesis & \makecell{sparsity \&\\learned dict.} & \makecell{transform coefs \&\\dictionary}  & $\ell_0$-min &  approximate & approximate & alternate GD \\
	\rowcolor{white!95!black}
  Chantas'18 \cite{Chantas2018:Inpainting_Variational_Bayesian_Inference} & synthesis & sparsity & transform coefs  & KL divergence & no & no & variational Bayes\\
  Gaultier'19 \cite{Gaultier2019:PhD.Thesis} & \makecell{analysis \&\\synthesis}& sparsity & transform coefs & $\ell_0$-min & yes & yes & ADMM \\
\rowcolor{white!95!black}
  \evaluated{Záviška'19} \cite{ZaviskaRajmicMokryPrusa2019:SSPADE_ICASSP} & synthesis & sparsity & transform coefs & $\ell_0$-min & yes & yes & ADMM \\ 
  %
 \evaluated{Záviška'19b} \cite{ZaviskaRajmicSchimmel2019:Psychoacoustics.l1.declipping} & synthesis & \makecell{sparsity \&\\ psychoacoust.}  & transform coefs & $\ell_1$-min & yes & yes & DR \\
  \hline
\end{tabular}

\vspace*{3pt}

Abbreviations:
ADMM: Alternating Direction Method of Multipliers,
AR: Autoregressive,
CV: Condat--V\~{u} algorithm,
CVX: convex opt.\ toolbox~\cite{CVX},\\
DR: Douglas--Rachford alg.,
EM: Expectation--Maximization,
(F)ISTA: (Fast) Iterative Shrinkage Thresholding Alg.,
GD: Gradient Descent,\\
IHT: Iterative Hard Thresholding,
KL: Kullback--Liebler,
LS: Least Squares,
ML: Maximum Likelihood,
NMF: Nonnegative Matrix Factorization,\\
(O)MP: (Orthogonal) Matching Pursuit,
RVP: Recursive Vector Projection,
SD: Semidefinite programming
\label{tab::methods_summary}
\end{center}
\end{table*}

\textbf{Roadmap of the article.}
Section \ref{sec:problem_formulation} formulates the problem and prepares the notation used in further parts.
A survey of declipping methods is given in Section~\ref{sec:Declipping.methods},
while the methods selected for a~thorough evaluation are described in more detail in Section~\ref{sec:Declipping.methods.selection}.
Then, the experiment and evaluation setup are explained in Section~\ref{sec:Evaluation}, together with the results and their discussion.
Finally, a conclusion is given and some perspectives for future research are indicated.

\section{Problem formulation}
\label{sec:problem_formulation}


In agreement with the clipping model \eqref{eq:clipping}, it is possible to divide the signal samples into three disjoint sets of indexes
$R,H,L$ such that $R\cup H\cup L = \{1,\ldots,N\}$ and, correspondingly,
to distinguish the \emph{reliable} samples (not influenced by clipping), samples \emph{clipped from above} to the high clipping threshold $\tc$ and samples \emph{clipped from below} to the low clipping threshold ($-\tc$), respectively.
The respective projection operators $\Mr$, $\Mh$ and $\Ml$ (masks) are used to select samples from the corresponding set.
%
With the mask operators, the following feasible sets can be defined:
%
%
\begin{subequations}
	\label{eq:gamma.subsets}
    \begin{align}
    	\Gammar &= 
    	\{\tilde{\x} 
    	\hspace{0.25em}|\hspace{0.25em} 
    	\Mr\tilde{\x} \hspace{-1.5pt}  =   \hspace{-1.5pt} \Mr\y
    	\}, \\
    	\Gammah &= 
    	\{\tilde{\x} 
    	\hspace{0.25em}|\hspace{0.25em} \Mh\tilde{\x} \hspace{-1.5pt}\geq  \hspace{-1.5pt} \tc\},\quad 
    	\Gammal = 
    	\{\tilde{\x} 
    	\hspace{0.25em}|\hspace{0.25em} \Ml\tilde{\x} \hspace{-1.5pt}\leq  \hspace{-1.5pt} -\tc\},\\
    	\Gamma &= \Gammar \cap \Gammah \cap \Gammal,
    \end{align}
\end{subequations}
\edt{where we use the tilde notation $\tilde{\x}$ to represent arbitrary time-domain signals,
while avoiding the confusion with $\x$, the clean signal.}
Note that these sets depend on the observation $\y$, since the masks do so too,
hence formally we should write $\Gamma(\y)$, for example,
but we omit the dependence on the signal at most places for brevity.

\edt{The original dynamic range of the signal before clipping is typically unknown.
However, if it is known or can at least be estimated,
additional constraints like $\Mh\tilde{\x} \leq \theta_\mathrm{max}$ and $\Ml\tilde{\x} \geq -\theta_\mathrm{max}$
 can be appended to \eqref{eq:gamma.subsets} to further restrict the feasible set $\Gamma$.
The scalars $\theta_\mathrm{min}$ and $\theta_\mathrm{max}$ represent the lower and upper bounds for the 
value of the signal.
}
For example,
\cite{Adler2011:Declipping} reports an improvement in signal recovery after such a~trick for heavily clipped signals.

The declipping task is clearly ill-posed, since there is an infinite number of solutions that satisfy \eqref{eq:gamma.subsets}.
Therefore, considering some additional information about the signal is crucial.
That is where a signal model or statistical model comes into play, which regularizes the inverse problem. 

\subsection{Preliminaries and some notation}
\label{sec:Preliminaries}
Most of the declipping methods rely on signal processing in a~transformed domain.
In such a context, \(A\colon\RR^N \to \todo{} \CC^P\) will denote the analysis operator,
and \(D\colon \CC^P \to \RR^N\) the synthesis operator.
The operators are linear, it holds $P\geq N$, and the operators are connected through the relation \(D = A^*\).
The asterisk $^*$ denotes the adjoint operator.
\edt{As an example, the common Discrete Fourier Transform (DFT) can play the role of $A$,
while $D$ is the inverse DFT.
The DFT belongs to the class of unitary transfroms, where it holds \(D = A^{-1}\).}

\edt{Time-frequency audio processing, however, benefits from cases where $P>N$.}
For computational reasons, authors often restrict themselves to
\edt{transforms referred to as the Parseval tight frames \cite{christensen2008},}
i.e., operators for which \(DD^* = A^*\!A = \Id\).
\edt{Here \(\Id\) stands for the identity operator.}
Unitary operators are clearly special cases of Parseval tight frames. 

In synthesis-based signal models, one seeks for coefficients $\z\in\CC^P$
that follow some prescribed properties, both directly in $\CC^P$
and after synthesizing the coefficients into the signal $D\z\in\RR^N$.
In the analysis models, one seeks for a~signal $\x\in\RR^N$ that satisfies some properties both in $\RR^N$ and after the analysis of $\x$ into coefficients, $A\x\in\CC^P$
\cite{Elad05analysisversus}.

In finite-dimensional spaces (which is our case), the operators $D$ and $A$ can be identified with matrices.
The matrix $D$ corresponding to the synthesis is often called \emph{the dictionary},
since its columns are the basic blocks in building the signal via their linear combination \cite{Bruckstein.etc.2009.SIAMReviewArticle}.

Since the methods covered by this survey concern exclusively audio,
it is natural that the majority of the methods use transforms
that map the signal to the time-frequency (TF) plane (and vice versa),
such as the short-time Fourier transform (STFT), often referred to as the discrete Gabor transform (DGT)
\cite{christensen2008,LTFAT}.
Methods based on such time-frequency transforms
work with (possibly overlapping) blocks of the signal.
Such signal chunks are usually created by means of time-domain windowing;
note that this is the reason why we will speak about signal \emph{windows}, alternatively about signal \emph{blocks},
but not about the time \emph{frames} of the signal,
in order to avoid confusion with the above-introduced concept of frames
in vector spaces.
The TF coefficients are treated in the form of the vector $\z$ from the mathematical point of view, 
but note that for the user it is often more convenient to form a~matrix $[z_{ft}]$ from $\z$.
Its rows represent frequencies and its columns correspond to time-shifts of the windows.
\edt{Such an arrangement corresponds to how spectrograms of audiosignals are usually visualized.}
The methods in Sections~\ref{ssec::dict_learn} and \ref{sec::nmf_declipping} will need to explicitly refer to
individual signal blocks.
For this sake, an additional index $t$ will be used, such that,
for instance, $\y_t$ will denote the $t$-th block of the clipped signal;
in analogy to this, appending $t$\/ to the masking operators and to the feasible sets will refer to their restriction in time, such as 
${\Mr}_t$ or $\Gamma(\y_t)=\Gamma_t$.

Norms of vectors will be denoted by $\norm{\cdot}$,
usually appended with the lower index characterizing the particular type of the norm.
Using no index corresponds to the case of the operator norm (i.e., the largest singular value of the operator).

Many methods are based on a~concept known
as `sparsity', popularized in the last two decades \cite{Bruckstein.etc.2009.SIAMReviewArticle,eladbook}.
Exploiting sparsity means that within feasible declipping solutions,
signals $D\z$ with a~low number of nonzero elements of $\z$ are prioritized (synthesis model)
or signals $\x$ with a~low number of nonzeros in $A\x$ are preferred (analysis model) \cite{eladbook}.
Mathematically, the $\norm{\cdot}_0$ pseudo-norm is used to count the nonzeros of a~vector.

\edt{Several methods presented}
below utilize the convex optimization;
typically a~sum of convex functions has to be minimized.
\edt{In line with the recent trend in convex optimization,}
numerical solutions of such problems will be found using the so-called \emph{proximal splitting algorithms} \cite{combettes2011proximal,Condat2014:Generic.proximal.algorithm,FadiliStarck2009:Monotone.op.splitting}.
These
are iterative schemes, usually with only a few principal---but rather simple---computations in each iteration.
Each such computational step is related to the respective function in the sum separately.
We recall that the proximal operator of a~convex function $f$ is the mapping
\begin{equation}
    \label{eq:Prox.definition}
    \prox_f (\x) = \argmin_{\z} \frac{1}{2} \norm{\x-\z}_2^2 + f(\z).
\end{equation}
The concept provides a generalization of the common projection operator \cite{combettes2011proximal}.

\subsection{\edt{Consistency of declipping and relation to inpainting}}
Whichever technique is employed, every declipping method can be assigned to one of several classes, 
based on
what is called
consistency.
A \emph{fully consistent} method
seeks a~solution that is a member of the
intersection
$\Gamma = \Gammar \cap \Gammah \cap \Gammal$, or in other words,
the recovered signal should equal the original samples in the reliable part
and, at the same time, it should lie beyond the clipping thresholds in the clipped part.
A~solution \emph{consistent in the reliable part} belongs to $\Gammar$,
while a~solution \emph{consistent in the clipped part} is a~member of $\Gammah \cap \Gammal$.
A~\emph{fully inconsistent method} does not require a~strict membership of the solution in any of the sets $\Gammar, \Gammah, \Gammal$.
\edt{Even in such a case, a~method either reflects the sets $\Gammah$ and $\Gammal$ (and thus the thresholds $\tc$ and $-\tc$ are taken into account in declipping) or these sets are completely ignored.}
\edt{Actually, if a~user (or a~particular application)
decides not to take $\Gammah$ and $\Gammal$ into consideration,
the declipping problem boils down to the so-called \emph{audio inpainting} problem.
Inpainting treats clipped samples simply as \emph{missing},
hence ignoring potentially useful information;
indeed, some of the presented methods in this survey are of this kind.
Audio inpainting itself is an area, see for example
\cite{MokryZaviskaRajmicVesely2019:SPAIN,Bahat_2015:Self.content.based.audio.inpaint,
Marafioti2020:GACELA}
and the references therein.
In our declipping experiment below, we included Janssen's method
\cite{javevr86}
as the representative of such methods
(being actually very successful in audio inpainting).}

Fig.\,\ref{fig:waveforms.consistency} shows examples of different types of solution consistency.
%
%
%
Consistent methods reflect the observed clipped signal and the clipping model, but the solutions are usually quite slow to compute.
Inconsistency usually means a gain in speed in exchange for only approximate solutions (which might still be great for the human auditory system).

\section{\mbox{\edt{Declipping methods not selected for evaluation}}}%
\vspace*{.5ex}
\label{sec:Declipping.methods}



Apart from the detailed treatment of the methods selected for further numerical  evaluation, this section is devoted to surveying the other declipping methods in the literature.
\edt{We remind the reader that the characterization of all declipping methods is summarized in Table~\ref{tab::methods_summary}.}


Abel and Smith
\cite{abel91_declipping}
discuss declipping of signals whose spectral band is limited (more than at the Nyquist frequency).
This assumption is the key ingredient leading to a convex optimization program.
The recovery uses oversampling and interpolation with sinc functions.
The method is fully consistent, apart from the \qm{noisy} variant treated at the end of~\cite{abel91_declipping}. 

Fong and Godsill \cite{FongGodsill2001:MonteCarlo}
approach the declipping problem from the viewpoint of Bayesian statistical signal processing.
The main assumption is the autoregressive nature of the signal,
and to find the declipped samples, Monte Carlo particle filtering is utilized.
The experiment follows a very simplified scenario
(a single test on a very short speech sample).

Dahimene {\it et al.}\ \cite{Dahimene2008_declipping}
also start from the autoregressive (AR) assumption imposed on the signal.
The paper forms a system of linear equations which is row-wise pruned in correspondence to the positions of the clipped samples.
Two means of signal estimation are suggested: one based on ordinary least squares and the other based on Kalman filtering.
This modeling does not guarantee any consistency in the clipped part.

The method introduced by Miura {\it et al.}\ \cite{Miura2011:DeclippingRecursiveVectorProjection}
is based on a~procedure coined recursive vector projection (RVP) by the authors.
It turns out that RVP is actually the classical matching pursuit algorithm
\cite{mallatzhang93}
restricted to reliable samples of the signal.
Thus, it is a~synthesis approach, with the dictionary described as the (possibly overcomplete) DFT.
Since the clipping constraints are not taken into consideration, 
\cite{Miura2011:DeclippingRecursiveVectorProjection}
is a~method inconsistent in the clipped part,
and its idea is actually quite similar to
filling the missing samples of audio
using the orthogonal matching pursuit in \cite{Adler2012:Audio.inpainting}.
 
Jonscher {\it et al.}\ \cite{Jonscher2014:DeclippingFSE} introduce the Frequency Selective Extrapolation method.
The signal is processed block by block.
The clipped samples are treated as missing and their reconstruction is performed sequentially.
The model behind the method can be understood as synthesis sparsity-based, with an overcomplete Fourier dictionary.
Tests were carried out on speech signals only.

The method proposed by Takahashi {\it et al.}\
\cite{Takahashi2013:Hankel_matrix_declipping, Takahashi2015:Block.Adaptive.Decipping.Null.Space}
starts from an interesting observation that 
when the Hankel matrix
is formed from a~signal that follows the autoregressive (AR) model,
the rank of this matrix is identical to the order of the respective AR process.
Therefore, the approach aims at estimating the unknown but clipped elements of the Hankel matrix, whose rank is being minimized at the same time.
After a series of simplifications, the authors formulate a~convex optimization problem.
The reported results look promising but unfortunately no data or codes are available.

Harvilla and Stern \cite{HarvillaStern2015:RBAR}
introduce the RBAR (Regularized Blind Amplitude Reconstruction) method,
which is reported to run in real time.
The declipping task is formulated as an extended Tikhonov-regularized least squares problem,
where the main idea is to penalize large deviations in the second difference of the signal,
and at the same time to penalize deviations from the clipping level in the clipped part.
The experiments were carried out on speech signals, no codes are available.

\edt{%
Selesnick \cite[Sec.\,8.6]{Selesnick2013:LeastSquares} uses a~similar regularizer.
He proposes to minimize the energy of the the third difference of the signal,
encouraging the filled-in data to have the form of a~parabola.
The numerical program is a~least squares problem and as such it has a closed-form solution.
Selesnick's method is however presented as one of the examples of the utilization of least squares and the paper does not provide any comparison with other methods.
}

\edt{%
Chantas {\it et al.} 
\cite{Chantas2018:Inpainting_Variational_Bayesian_Inference}
build their declipping algorithm upon the Bayesian inference.
They use a~synthesis model, where the discrete cosine transform (DCT) coefficients are modeled as following the Student's distribution, complying with the assumption of their sparsity.
No utilization of clipping constraints is involved.
In addition, no computer implementation is available, unfortunately.}

Elvander {\it et al.} \cite{Elvander2017:Gridless.estimation.saturated.signals}
introduce probably the first approach that adapts the grid-less sparse recovery framework to the declipping problem. 
In brief, grid-less means that the dictionary does no longer contain countably many columns.
In the case of \cite{Elvander2017:Gridless.estimation.saturated.signals}, the continuous range of frequencies is available as the building blocks for the signal.
A~drawback of such an approach is that
the resulting minimization problem is a~semidefinite program, which can be computationally expensive.

In his PhD thesis, Gaultier \cite{Gaultier2019:PhD.Thesis} extends the idea of earlier algorithms based on hard thresholding
\cite{Kitic2013:Consistent.iter.hard.thresholding,KiticBertinGribonval2014:AudioDeclippingCosparseHardThresholding,Kitic2015:Sparsity.cosparsity.declipping}.
The author works with the idea similar to the one published in 
\cite{SiedenburgKowalskiDoerfler2014:Audio.declip.social.sparsity},
and introduces coefficients neighborhoods such that the TF coefficients are not processed individually (as is commonly done) but group-wise.
This is shown to be beneficial, especially for mild clipping.
A~comprehensive study based of such an approach is contained in a~recent article
\cite{GaultierKiticGribonvalBertin:Declipping2020.preprint}.

This survey considers only hard clipping governed by
Eq.\,\eqref{eq:clipping}
but 
let us shortly mention the existence of the soft clipping (and the corresponding declipping methods).
The transfer function of the soft clipping does not break suddenly at the points $-\tc$ and $\tc$ as in the case of hard clipping.
Rather a~certain transition interval is present around these spots that makes the transfer function smoother,
resulting in less spectral distortion of clipping.
The recovery of signals that have been soft-clipped is treated in
\cite{Avila2017:Soft.Declipping.Least.Squares},
\cite{Avila2017:Soft.Declipping.weighted.l1},
\cite{GorlowReiss2013:Inversion.DR.Compression},
to name but a~few.




\section{Declipping methods selected for evaluation}
\label{sec:Declipping.methods.selection}

This section explains the principles of the methods that have been selected for the evaluation procedure.
Each method comes with the algorithm in pseudocode
(software implementation is addressed later in Section~\ref{sec:Evaluation}).
%
Some of the existing methods based on the synthesis sparsity are quite easily adaptable to the respective analysis counterpart;
we include such unpublished variants in several cases to cover a~wider range of methods.
The order of the methods in this Section is chosen such that
\edt{the $\ell_0$-based are covered first},
then the $\ell_1$-based (optionally including psychoacoustics) are presented,
then methods that can adapt to the signal,
and as the last one the simple Janssen method \cite{javevr86} serving as the \qm{anchor}.

\subsection{Constrained Orthogonal Matching Pursuit (C-OMP)}
%
The approach to audio declipping proposed by Adler {\it et al.} \cite{Adler2011:Declipping} follows the same idea as the article \cite{Adler2012:Audio.inpainting} devoted to inpainting.
The Orthogonal Matching Pursuit (OMP) is a~well-known
greedy-type algorithm
\cite{tropp_greed,SturmChristensen2012:Comparison.of.OMP}
used here as the first part of the procedure that approximates sparse coefficients in the NP-hard problem
\begin{equation}
    \label{eqn:cOMP}
    \argmin_{\z} \norm{\z}_0 \ \, \text{s.t.}\
    \begin{cases}
        \norm{\Mr\y-\Mr D\z}_2 \leq\epsilon,\\
        D\z \in \Gammah \cap \Gammal.
    \end{cases}
\end{equation}
%
It is clear that \eqref{eqn:cOMP} is a synthesis-based signal model and that it is clip-consistent,
but inconsistent in the reliable part.
The authors of \cite{Adler2011:Declipping} use an overcomplete
discrete cosine transform
in the role of the synthesis operator $D$.

The signal is cut into overlapping windows first,
and the OMP is applied in each window separately.
In the course of the OMP iterations,
an increasing number of
significant coefficients with respect to $D$ are picked in a~greedy fashion.
Once such a~subset of coefficients fulfils
$\norm{\Mr\y-\Mr D\z}_2 \leq\epsilon$,
where $\epsilon>0$ is the user-set precision required for the reliable part,
the OMP stops.
Notice that doing this is effectively performing the audio inpainting using OMP,
\edt{i.e., ignoring the second condition in~\eqref{eqn:cOMP}}.
However, as the very last step of the estimation,
the current solution is updated using convex optimization.
This makes the approach very slow,
since it requires an iterative algorithm.
The authors of \cite{Adler2011:Declipping} rely in this step on the CVX toolbox \cite{CVX} in which (the subset of) $D$ is handled in the matrix form, which deccelerates the computations even more.
%
After the algorithm is finished, the coefficients are synthesized using $D$.
Individual blocks of the declipped signal are then put together in the time domain, using the common overlap--add procedure.
The algorithm for a~single window is summarized in Alg.\,\ref{alg:Constrained.OMP}.

\begin{algorithm}
    \setstretch{1.1}
    \DontPrintSemicolon
    \SetAlgoVlined
    \SetKwInput{Input}{Input}
    \SetKwInput{Par}{Parameters}
    \SetKwInput{Init}{Initialization}
	\Input{$D,\ \y\in\RR^N,\ R,\ H,\ L$}
 	\Par{$\epsilon > 0$}
	%
	\setstretch{1.2}
	Using OMP, find an approximate solution, $\hat{\z}$, to problem\linebreak
	\hspace*{1.5em}%
	$\argmin_{\z} \norm{\z}_0 \ \, \text{s.t.}\ \norm{\Mr\y-\Mr D_{\w}\z}_2 \leq\epsilon$\;
    Fix the support $\Omega\subseteq\{1,\ldots,P\}$ of $\hat{\z}$\;
    Solve the constrained convex program\linebreak%
    \hspace*{-0.5em}%
    $\hat{\z}_\Omega =
    \argmin_{\z_\Omega} \norm{\Mr(\y-D_\Omega \z_\Omega)}_2\
		\text{s.t.}\ D_\Omega \z_\Omega \in \Gammah \cap \Gammal$\;
	\KwRet{$D_\Omega \hat{\z}_\Omega$}
	\caption{Constrained OMP declipping \cite{Adler2011:Declipping} (C-OMP)}
	\label{alg:Constrained.OMP}
\end{algorithm}

Some remarks should be made here:
First, consider $D$ as a~matrix for the moment; the OMP requires that the columns of $D$ have the same energy,
i.e., the same $\ell_2$ norm---this kind of normalization guarantees a~fair selection of coefficients,
at least for oscillatory signals, such as audio.
To preserve such a~condition, the problem on line~1 of Alg.\,\ref{alg:Constrained.OMP} needs to weight the columns of $D$,
which arises from the fact that the \edt{column subvectors}
used for estimation, $\Mr D$,
do not contain the same energy.
We denote this weighted synthesis $D_{\w}$.
For more details, see the original paper \cite{Adler2011:Declipping} or
the discussion on different types of weighting in \cite{MokryRajmic2020:Inpainting.revisited}.

Second, Alg.\,\ref{alg:Constrained.OMP} uses the notation $D_\Omega$ for the synthesis operator restricted just to its columns contained in the set $\Omega$,
and, by analogy, for the restricted vector of coefficients $\z_\Omega$.
There is in fact no need to weight the columns of $D_\Omega$ for the purpose of solving the problem on line 3.

Third, notice that the condition
$\norm{\Mr\y-\Mr D\z}_2 \leq\epsilon$
is in general violated after the update at line 3
because there might be no solution to the convex program that would satisfy it.
Furthermore,
hand in hand with this,
notice that while $D$ is usually chosen as a~frame in $\RR^N$,
the restriction $D_\Omega$ does not have to inherit this property anymore,
and this fact naturally applies to $\Mr D_\Omega$ as well.
In turn, when OMP finds $\Omega\subset\{1,\ldots,P\}$,
there is
no guarantee of the existence of \emph{any}\/ solution to the convex program.
We will return to this issue in the experimental part
since it will serve as an explanation of the strange numerical behavior of the C-OMP in some rare cases.

\subsection{A-SPADE}
\label{subsec:aspade}
%
The A-SPADE (Analysis SPArse DEclipper) was introduced by Kiti\'c {\it et al.} in \cite{Kitic2015:Sparsity.cosparsity.declipping}.
It is a~natural successor to similar sparsity-based approaches
\cite{Kitic2013:Consistent.iter.hard.thresholding} and \cite{KiticBertinGribonval2014:AudioDeclippingCosparseHardThresholding},
which are outperformed by A-SPADE \cite{Kitic2015:Sparsity.cosparsity.declipping}.

The A-SPADE algorithm 
\edt{belongs to the $\ell_0$-based family}
and it approximates the solution of the following NP-hard
regularized inverse problem
\begin{equation}
 	\label{eq:SPADE.analysis.formulation}
	\min_{\x,\z} \norm{\z}_0 \ \, \text{s.t.}
	\ \,\x\in\Gamma(\y) \ \text{and}\ 
	\norm{A\x-\z}_2 \leq\epsilon.
\end{equation}
Here,
\(\x \in \RR^N\) stands for the unknown signal in the time domain,
and \(\z \in \CC^P\) contains the (also unknown) coefficients.

Parseval tight frames \cite{christensen2008}, i.e., \(DD^* = A^*\!A = \Id\), are considered.
The processing of the signal is
sequential,
window by window.
Due to the overlaps of windows, it suffices to use a simple TF transform like the DFT or the DCT (both possibly oversampled) which are Parseval tight frames.

The optimal solution to \eqref{eq:SPADE.analysis.formulation} in each window is  approximated by means of the alternating direction method of mutipliers (ADMM).
The resulting algorithm is in Alg.\,\ref{alg:aspade};
for a~detailed derivation and discussion, see \cite{ZaviskaMokryRajmic2018:SPADE_DetailedStudy}.
The computational cost of the A-SPADE is dominated by the signal transformations;
the algorithm requires one synthesis and one analysis per iteration.
The hard thresholding $\mathcal{H}$ is a trivial operation.
The projection on line~3 seeks for the signal $\x\in\Gamma$ whose analysis $A\x$ is the closest to 
$\bar{\z}^{(i+1)}-\u^{(i)}$.
For tight frames, this task can be translated to an elementwise mapping in the time domain
\cite{ZaviskaMokryRajmic2018:SPADE_DetailedStudy},
%
\begin{equation}
\Big(\proj_{\Gamma}(\u)\Big)_n = \left\{
	\begin{array}{ll}
		y_n & \text{for } n \in R, \\
		\max(\tc, u_n) & \text{for } n \in H, \\
		\min(-\tc, u_n) & \text{for } n \in L,
	\end{array}
	\right.
\label{eq:proj_time_short}
\end{equation}
$(\u)_n$ denoting the $n$th element of the vector, i.e., $(\u)_n=u_n$.

Compared with most available algorithms, it is fairly easy to tune the parameters.
The variable $k$ directly represents the number of selected coefficients in the hard-thresholding step.
This number is growing in the course of iterations, driven by the parameters $s$ and $r$
(every $r$-th iteration, $k$ is increased by $s$).
The algorithm works really well 
with the basic setting, where $k$ increases by one (i.e., $s=r=1$).

	

\begin{algorithm}
    \setstretch{1.1}
    \DontPrintSemicolon
    \SetAlgoVlined
    \SetKwInput{Input}{Input}
    \SetKwInput{Par}{Parameters}
    \SetKwInput{Init}{Initialization}
	\Input{$A,\ \y\in\RR^N,\ R,\ H,\ L,\ \epsilon > 0$}
	\Par{$s,\,r \in \NN$} 
	\Init{$\hat{\x}^{(0)}\in\RR^N,\ \u^{(0)}\in\CC^P,\ k = s$}
	\setstretch{1.2}
	\For{$i=0,1,\dots$\,\upshape\textbf{until} $\norm{A\x-\z}_2 \leq\epsilon$ }{
	    \(\bar{\z}^{(i+1)} = \mathcal{H}_k\left(A\hat{\x}^{(i)}+\u^{(i)}\right)\)\;
	    \({\hat{\x}^{(i+1)} \hspace{-1pt}=\hspace{-1pt} \argmin_\x\hspace{-1pt}{\|A\x\hspace{-1pt}-\hspace{-1pt}\bar{\z}^{(i+1)}\hspace{-1pt}+\hspace{-1pt}\u^{(i)}\|_2^2}}
        \text{\hspace{0.5em}s.t.\hspace{0.5em}}\x \in \Gamma\)\;
	    \(\u^{(i+1)}=\u^{(i)}+A\hat{\x}^{(i+1)}-\bar{\z}^{(i+1)}\)\;
    	\lIf{$(i+1)\modulo r = 0$}{\(k = k+s\)}
	}
	\KwRet{$\hat{\x}^{(i+1)}$}
	\caption{\mbox{A-SPADE algorithm \cite{Kitic2015:Sparsity.cosparsity.declipping}}%
	}
	\label{alg:aspade}
\end{algorithm}

\subsection{S-SPADE} 
\label{subsec:sspade}
%
Similarly to A-SPADE, Kiti\'c {\it et al.} \cite{Kitic2015:Sparsity.cosparsity.declipping} also introduce a~syn\-thesis-based formulation,
\begin{equation}
 	\label{eq:SPADE.synthesis.formulation}
	\min_{\x,\z} \|\z\|_0 \ \, \text{s.t.}
	\ \,\x\in\Gamma(\y) \ \text{and}\ 	
	\norm{\x-D\z}_2 \leq\epsilon.
\end{equation}
However, the S-SPADE algorithm from \cite{Kitic2015:Sparsity.cosparsity.declipping}
has been shown in 
\cite{ZaviskaRajmicMokryPrusa2019:SSPADE_ICASSP,ZaviskaMokryRajmic2018:SPADE_DetailedStudy}
as actually  solving a~different optimization problem than \eqref{eq:SPADE.synthesis.formulation}.
The same publications suggested a~new version of the S-SPADE as the true counterpart of the A-SPADE,
and showed its superiority in
performance.
Such an algorithm is in Alg.\,\ref{alg:sspade_dr}.

The computational complexity of Algs.\ \ref{alg:aspade} and \ref{alg:sspade_dr} is the same:
one synthesis, one analysis, the hard thresholding and an elementwise mapping per iteration is employed.


\begin{algorithm}
    \setstretch{1.1}
    \DontPrintSemicolon
    \SetAlgoVlined
    \SetKwInput{Input}{Input}
    \SetKwInput{Par}{Parameters}
    \SetKwInput{Init}{Initialization}
    %
	\Input{$D,\ \y\in\RR^N,\ R,\ H,\ L,\ \epsilon > 0$}
	\Par{$s,\,r \in \NN$} 
	\Init{$\hat{\x}^{(0)}\in\RR^N,\ \u^{(0)}\in\RR^N,\ k = s$}
	\setstretch{1.2}
	\For{$i=0,1,\dots$\,\upshape\textbf{until} $\norm{\x-D\z}_2 \leq\epsilon$}{ 
	\(\bar{\z}^{(i+1)} = \mathcal{H}_k\left(D^*(\hat{\x}^{(i)} - \u^{(i)})\right)\)\;
	\({\hat{\x}^{(i+1)} \hspace{-1pt} = \argmin_\x{ \hspace{-2pt} \|D\bar{\z}^{(i+1)}\hspace{-1pt}-\hspace{-1pt}\x\hspace{-1pt}+\hspace{-1pt}\u^{(i)} \hspace{-1pt} \|_2^2}}
	\hspace{0.5em}\text{s.t.}\hspace{0.5em}\x \in \Gamma\)\hspace{-1em}\;
	\(\u^{(i+1)}=\u^{(i)}+D\bar{\z}^{(i+1)}-\hat{\x}^{(i+1)}\)\;
	    \lIf{$(i+1)\modulo r = 0$}{\(k = k+s\)}
	}
	\KwRet{\(\hat{\x}^{(i+1)}\)}
	\caption{\mbox{S-SPADE algorithm according to \cite{ZaviskaRajmicMokryPrusa2019:SSPADE_ICASSP}}
	}
	\label{alg:sspade_dr}
\end{algorithm}

\subsection[Declipping using weighted l1 minimization]{Declipping using weighted $\ell_1$ minimization}
\label{sec:Simple.weighted.l1.approach}


The above methods are based \edt{on the $\ell_0$ approach.} 
Now we present several methods that rely on what is known as
convex relaxation:
their idea is to substitute the non-convex $\ell_0$ pseudonorm with the \qm{closest} convex norm, which is the $\ell_1$
\cite{Bruckstein.etc.2009.SIAMReviewArticle,DonohoElad2003:Optimally}.
The two declipping formulations in this section are quite basic,
but to our knowledge they are treated only in
\cite{RajmicZaviskaVeselyMokry2019:Axioms}
and 
\cite{ZaviskaRajmicSchimmel2019:Psychoacoustics.l1.declipping}.
In this article, they are included in the evaluation in their simple form but they also serve as the building block for
algorithms from further sections.

Let us start with the simple synthesis-based task  
\begin{equation}
    \argmin_{\z} \norm{\w\odot\z}_1 \text{\ s.t.\ } D\z \in \Gamma, 
    \label{eq:problem_syn.basic}
\end{equation}
where $D$
is the synthesis operator 
and $\odot$ denotes the elementwise product of two vectors.
The vector $\w>0$
\edt{is the vector of weights that}
can be set to all ones when no coefficients should be prioritized,
but the larger an element of $\w$ is, the more penalized the corresponding coefficient in $\z$ is in the optimization.
Due to the strict requirement $D\z \in \Gamma$, the approach based on \eqref{eq:problem_syn.basic} is fully consistent.

To find an appropriate algorithm to solve \eqref{eq:problem_syn.basic}, it is convenient to rewrite it in an unconstrained form: 
\begin{equation}
    \argmin_{\z} \norm{\w\odot\z}_1 + \iota_\Gamma(D\z), 
    \label{eq:problem_syn.basic.unconst}
\end{equation}
where the hard constraint from \eqref{eq:problem_syn.basic} has equivalently been replaced by the indicator function $\iota$,
%
\begin{equation}
\iota_C (\u) = \left\{
	\begin{array}{ll}
			0 & \text{for } \u \in C, \\
			+\infty & \text{for } \u \notin C.
	\end{array}
\right.
\label{eq:indicator_func}
\end{equation}
Next, the observation is used that $\iota_\Gamma(D\z) = (\iota_\Gamma \circ D)(\z) = \iota_{\Gamma^*}(\z)$,
with $\Gamma^*$ being the set of coefficients consistent with the clipping model,
i.e., $\Gamma^* = \{ \tilde{\z} \mid D\tilde{\z}\in \Gamma \}$;
see also the
definitions \eqref{eq:gamma.subsets}.
The Douglas--Rachford algorithm (DR) \cite{combettes2007douglas} is able to find the minimizer of a~sum of two convex functions of our type.
The algorithm is presented in Alg.\,\ref{alg:DR.declipping}.
%
%
\begin{algorithm}
    \setstretch{1.1}
	\DontPrintSemicolon
	\SetAlgoVlined
    \SetKwInput{Input}{Input}
    \SetKwInput{Par}{Parameters}
    \SetKwInput{Init}{Initialization}
	\Input{$D,\ \y\in\RR^N,\ \w\in\RR^P,\ R,\ H,\ L$}
	\Par{$\lambda = 1,\ \gamma > 0$}
	\Init{$\z^{(0)} \in \CC^P$}
	\setstretch{1.2}
	\For{$i=0,1,\dots$\,\upshape\textbf{until} convergence}{ 
		$\tilde{\z}^{(i)} = \proj_{\Gamma^*}\z^{(i)}$ \comment{using \eqref{eq:projection}} \;
		$\z^{(i+1)} = \z^{(i)} + \lambda \left(\soft_{\gamma\w}(2\tilde{\z}^{(i)} - \z^{(i)}) - \tilde{\z}^{(i)} \right)$ \;
	}
	\KwRet{$D\z^{(i+1)}$}
	\caption{%
	\mbox{Douglas--Rachford (DR) alg.\ solving
    \eqref{eq:problem_syn.basic}
    \cite{RajmicZaviskaVeselyMokry2019:Axioms}}
	}
	\label{alg:DR.declipping}
\end{algorithm}

The algorithm iterates over two principal steps:
The first is the projection onto $\Gamma^*$, which corresponds to the proximal operator of 
$\iota_{\Gamma^*}$
(recall the definition of prox in \eqref{eq:Prox.definition}).
This projection is nontrivial and an explicit formula exists only in the case when 
\(DD^*\) is diagonal.
According to \cite{RajmicZaviskaVeselyMokry2019:Axioms},
for Parseval tight frames it holds
\begin{equation}
    \proj_{\Gamma^*}(\z) = \z - D^*\left(D\z-\proj_{\Gamma}(D\z)\right),
    \label{eq:projection}
\end{equation}
where $\proj_{\Gamma}$
is a~trivial, elementwise time-domain mapping.
%
The second step involves the soft thresholding $\soft_{\tau\w}$ with the vector of thresholds $\tau\w$, which coincides with the proximal operator of the weighted $\ell_1$-norm. 
The soft thresholding is an elementwise mapping defined by 
\begin{equation}
\soft_{\tau\w}(\z) = \z \odot
        \max \left(1-\tau\w\odot \frac{1}{\abs{\z}},0 \right).    
\end{equation}

The analysis counterpart of \eqref{eq:problem_syn.basic} reads  
\begin{equation}
    \argmin_{\x} \norm{\w\odot A\x}_1 \text{\ s.t.\ } \x \in \Gamma, 
    \label{eq:problem_ana.basic}
\end{equation}
where $A$
is the analysis operator.
Unfortunately, the presence of $A$ inside the weighted $\ell_1$ norm
prevents using the DR algorithm as above.
The Chambolle--Pock (CP) algorithm \cite{ChambollePock2011:First-Order.Primal-Dual.Algorithm} is able to cope with problems including a general linear operator.
Its particular form for signal declipping is shown in Alg.\,\ref{alg:CP.for.declipping}.
%
%
\begin{algorithm}
    \setstretch{1.1}
    \DontPrintSemicolon
    \SetAlgoVlined
    \SetKwInput{Input}{Input}
    \SetKwInput{Par}{Parameters}
    \SetKwInput{Init}{Initialization}
    \Input{$A,\ \y\in\RR^N,\ \w\in\RR^P,\ R,\ H,\ L$}
    \Par{$\zeta,\, \sigma > 0$ and $\rho\in[0,1]$}
    \Init{$\x^{(0)} \in \RR^N,\ \vv^{(0)} \in \CC^P$}
    \setstretch{1.2}
    \For{$i=0,1,\dots$\,\upshape\textbf{until} convergence}{%
    $\vv^{(i+1)} = \clip_{\w}(\vv^{(i)}+\sigma \ana \bar{\x}^{(i)})$\\
    	$\x^{(i+1)} = \proj_{\Gamma}(\x^{(i)}-\zeta \ana^*{\vv}^{(i+1)})$\\
    	$\bar{\x}^{(i+1)} = \x^{(i+1)} + \rho (\x^{(i+1)} - \x^{(i)}) $\\
    }
    \KwRet{\(\bar{\x}^{(i+1)}\)}
    \caption{\mbox{Chambolle--Pock (CP) algorithm solving \eqref{eq:problem_ana.basic}}
    }
    \label{alg:CP.for.declipping}	
\end{algorithm}
There, $\clip$ is the Fenchel--Rockafellar conjugate of the soft thresholding,
defined as
\begin{equation}
    \clip_{\w}(\x) = (\Id - \soft_{\w})(\x).
    \label{eq:clip}
\end{equation}
For $\rho=1$, the CP algorithm converges if $\zeta\sigma\norm{A}^2<1$.

Looking at Algorithms \ref{alg:DR.declipping} and \ref{alg:CP.for.declipping}, one recognizes that both have identical cost per iteration.
The dominant operations are the transformations $A$ and $D$.


\subsection[Declipping in Sparseland (Rl1CC)]{Declipping in Sparseland (R$\ell_1\!$CC)}
%
Weinstein and Wakin \cite{Weinstein2011:DeclippingSparseland}
present four approaches to declipping, all based on sparsity.
The basic synthesis model \eqref{eq:problem_syn.basic} is actually covered by the article under the acronym BPCC (Basis Pursuit with Clipping Constraints).
%
In this section, we review the most successful method of \cite{Weinstein2011:DeclippingSparseland} with coefficient reweighting,
referred to as R$\ell_1\!$CC  (Reweighted $\ell_1$ with Clipping Constraints) by the authors.
It is again a~fully consistent approach.

R$\ell_1\!$CC follows up on a~well-known 
idea from the field of sparse recovery\,/\,compressed sensing:
To enhance the sparsity of the solution, a~standard iterative program is performed,
but repeatedly.
\edt{During the repetitions, the actual weights $\w$ are being adapted based on the current temporary solution
\cite{CandesWakinBoyd2008:enhanced.reweighted.l1}
and since each weight is inversely proportional to the magnitude of the respective coefficients,}
large coefficients are penalized less and less during the course of 
runs. 
For tiny coefficients, the opposite is true, leading to a sharper final sparsity and, in effect, to a~significantly better bias of the solution \cite{HastieTibshirani2015:Statistical.learning.LASSO,Rajmic2003:Exact.risk.analysis}.
The described effect, however, is not achieved automatically;
in some applications, the improvement can be large compared with the non-adaptive case \cite{MokryRajmic2019:Reweighted.l1.inpainting}, but sometimes it does not improve much \cite{NovosadovaRajmic2017:Piecewise.polynomial.segment.reweighted.optimiz}
or even fails.
It is worth noticing that it is not correct to say that R$\ell_1\!$CC solves \eqref{eq:problem_syn.basic}, see the discussion in \cite{CandesWakinBoyd2008:enhanced.reweighted.l1}.

To be more specific, R$\ell_1\!$CC starts by solving the problem \eqref{eq:problem_syn.basic} with weights set to $\w=\One$.
Based on the solution, $\w$ is recomputed and \eqref{eq:problem_syn.basic} is solved again, and again, until a~reasonable convergence criterion is fulfilled.
The authors of \cite{Weinstein2011:DeclippingSparseland}, however, provide no algorithm to solve \eqref{eq:problem_syn.basic}.\!%
\footnote{Codes
from \url{https://github.com/aweinstein/declipping} rely on
CVX \cite{CVX}.}
We know from Sec.\,\ref{sec:Simple.weighted.l1.approach} that the DR algorithm can be used.
In turn, R$\ell_1\!$CC is presented in Alg.\,\ref{alg:DR.declipping.reweighted}, with reweighting performed in step 4.
Note that in practice, the number of outer loop repetitions should be controlled in order to avoid drop in performance;
see the discussion in the evaluation part.


\begin{algorithm}
    \setstretch{1.1}
	\DontPrintSemicolon
	\SetAlgoVlined
	\SetKwInput{Input}{Input}
    \SetKwInput{Par}{Parameters}
    \SetKwInput{Init}{Initialization}
	\Input{$D,\ \y\in\RR^N,\ R,\ H,\ L$}
	\Par{$\epsilon>0$ }
	\Init{$\z^{(0)} \in \CC^P,\ \w^{(0)} = \mathbf{1}$}
	\setstretch{1.2}
    \For{$i=0,1,\dots$\,\upshape\textbf{until} convergence}{	
        Solve \eqref{eq:problem_syn.basic} using Alg.\,\ref{alg:DR.declipping} with $\w^{(i)}$ \comment{returns $\z^{(i+1)}$} \;
		$\w^{(i+1)}= \frac{1}{\abs{\z^{(i+1)}} + \epsilon}$  \comment{update weights elementwise}\;
	}
	\KwRet{$D\z^{(i+1)}$}
	\caption{R$\ell_1\!$CC using the Douglas--Rachford alg.
	}
	\label{alg:DR.declipping.reweighted}
\end{algorithm}


For this survey, we found it interesting to also include the analysis variant that was not considered in \cite{Weinstein2011:DeclippingSparseland}.
The procedure is analogous to the one just presented:
Problem \eqref{eq:problem_ana.basic} is solved repeatedly, now by the CP algorithm,
and the weights $\w$ are adapted depending on the last available solution,
similar to how it is done in Alg.\,\ref{alg:DR.declipping.reweighted}.
However, the difference is that the solution of the CP algorithm is a~time-domain signal;
recomputing the weights thus requires the application of an additional analysis operator.


\subsection{Social Sparsity}
\label{sec:social.sparsity}
%
Siedenburg {\it et al.} \cite{SiedenburgKowalskiDoerfler2014:Audio.declip.social.sparsity}
utilize for audio declipping the concept of the so-called `social sparsity' \cite{kowalski2012social,Bayram2011:Mixed.norms.overlapping.groups}.
The plain sparsity induced by the $\ell_1$ norm as the regularizer,
used in the above sections,
resulted in the soft thresholding of each coefficient individually in the respective algorithms.
The social sparsity approach is more general:
it allows the shrinkage of a~coefficient, based \edt{also on the values of other coefficients, typically the coefficients in a~kind of \qm{neighborhood}}.

The particular design of the neighborhood depends heavily on the task to solve.
For declipping, i.e., reverting a~time-domain damage, TF neighborhoods that correspond to spreading in the direction of time are beneficial, since they help to share and leak information in the time direction.
Such neighborhoods promote persistence in time, since with clipping, it makes more sense to focus on harmonic structures in audio than on transients.

Mathematically, the problem to solve is 
%
    \begin{multline}
    \label{eq:Social.sparsity.problem}
    \min_{\z}
    \left\{
        \frac{1}{2}\norm{\Mr D\z - \Mr\y }_2^2 
        + \frac{1}{2}\norm{h(\Mh D\z-\Mh\tc\One)}_2^2 + {} \right. \\
        \left. {} + \frac{1}{2}\norm{h(-\Ml D\z-\Ml\tc\One)}_2^2 + \lambda R(\z)
    \right\}.
    \end{multline}
%
It is a~synthesis model and it allows inconsistency of the reliable part (see the first term).
The terms with $h$ penalize the distance of the solution $D\z$ from the feasible set $\Gamma$ in the clipped part;
function $h$, called hinge, acts elementwise such that for each element of its input,
\begin{equation}
    h(u) =
    \begin{cases}
        u & \text{for}\ u<0\\
        0 & \text{otherwise.}
    \end{cases}
\end{equation}
The bold symbol $\One$ represents the vector of ones which is as long as the signal.
Since the use of $h$ in \eqref{eq:Social.sparsity.problem} does not guarantee that $D\z$ stays above and below the clipping thresholds,
the method \cite{SiedenburgKowalskiDoerfler2014:Audio.declip.social.sparsity}
is also inconsistent in the clipped part.

The first three terms in \eqref{eq:Social.sparsity.problem} are clearly differentiable,
even with the Lipschitz-continuous gradient.
Therefore, \eqref{eq:Social.sparsity.problem} can be treated as a sum of two functions,
the second of them, $R$, being possibly non-smooth.
This observation makes it possible to use standard optimization algorithms such as
ISTA or FISTA \cite{DaubechiesDefriseMol2004:ISTA,beck2009fast,combettes2011proximal}, as outlined in Alg.\,\ref{alg:social.sparsity}.


\begin{algorithm}
\setstretch{1.1}
\SetAlgoVlined
\DontPrintSemicolon
\SetKwInput{Input}{Input}
\SetKwInput{Init}{Initialization}
\SetKwInput{Par}{Parameters}
	\Input{$\y\in\RR^N,\ \lambda>0,\ R,\ H,\ L,\ D$; the shrinkage operator $\mathcal{S}$ (L\,/\,WGL\,/\,EW\,/\,PEW)}
	\Par{$\gamma\in\RR,\ \delta = \norm{DD^*}$}
	\Init{$\hat{\z}^{(0)},\, \z^{(0)} \in \CC^P$}
	\setstretch{1.2}
    \For{$i=0,1,\dots$\,\upshape\textbf{until} convergence}{ 	
    \(\g_1 = D^* \Mr^* (\Mr D\z^{(i)} - \Mr\y)\) \comment{gradients} \;
	\(\g_2 = D^* \Mh^*\, h(\Mh D\z^{(i)} - \Mh\tc\One) \)\;
	\(\g_3 = D^* \Ml^*\, h(-\Ml D\z^{(i)} - \Ml\tc\One) \)\;
	\(\hat\z^{(i+1)}\!=\mathcal{S}_{\lambda/\delta} \left( {\z}^{(i)}\! - \frac{1}{\delta} (\g_1+\g_2+\g_3) \right) \) \hspace{-1.1em}\mbox{\comment{step,\,shrink}}\;
	\( {\z}^{(i+1)} = \hat\z^{(i+1)} + \gamma \,( \hat\z^{(i+1)} - \hat\z^{(i)} ) \)  \comment{extrapolate} \; 
    }
 	\KwRet{$D\hat{\z}^{(i+1)}$}
	\caption{\mbox{ISTA-type Social sparsity declipper \cite{SiedenburgKowalskiDoerfler2014:Audio.declip.social.sparsity}}}
	\label{alg:social.sparsity}
\end{algorithm}

In step 5 of Alg.\,\ref{alg:social.sparsity}, the gradients are added.
Looking at the structure of the particular gradients on lines 2--4 reveals that in practical implementation,
a much more effective way of computing $\g_1+\g_2+\g_3$ is possible, containing a~single application of $D$ and $D^*$.
Another important trick that is not included in Alg.\,\ref{alg:social.sparsity} for clarity of presentation is the warm start/adaptive restart strategy \cite{DonoghueCandes2013:Adaptive.restart}:
the authors of \cite{SiedenburgKowalskiDoerfler2014:Audio.declip.social.sparsity} found out that the overall convergence is significantly accelerated
if ISTA is first run for a~large $\lambda$ for a~few hundred iterations,
then $\lambda$ is decreased and so on, until the target value of $\lambda$ from 
\eqref{eq:Social.sparsity.problem} is reached.

In Alg.\,\ref{alg:social.sparsity}, the shrinkage operator $\mathcal{S}$ plays the role of the proximal operator of $R$.
The regularizer $R$\/ should promote the expected structure of the TF coefficients of the signal.
Paper \cite{SiedenburgKowalskiDoerfler2014:Audio.declip.social.sparsity} suggests using four types of social shrinkage of the TF coefficients $\z$, indexed by $t$ (in time) and $f$ (in frequency):
%
\begin{itemize}
    \item
        LASSO (L):\quad
        $\mathcal{S}_\lambda(z_{ft}) = z_{ft} \cdot
        \max \left(1-\frac{\lambda}{\abs{z_{ft}}},0 \right)$.
    %
    \vspace{.6em}
    \item
        Windowed Group LASSO (WGL):\\
        $\mathcal{S}_\lambda(z_{ft}) = z_{ft} \cdot
        \max \left(1-\frac{\lambda}{\norm{\mathcal{N}(z_{ft})}},0 \right)$,\\
        where $\mathcal{N}(z_{ft})$ denotes a vector formed from coefficients in the neighborhood of TF position $ft$.
    \vspace{.6em}
    \item
        Empirical Wiener (EW):\\
        $\mathcal{S}_\lambda(z_{ft}) = z_{ft} \cdot
        \max \left(1-\frac{\lambda^2}{\abs{z_{ft}}^2},0 \right)$.
    \vspace{.6em}
    \item
        Persistent Empirical Wiener (PEW):\\
        $\mathcal{S}_\lambda(z_{ft}) = z_{ft} \cdot
        \max \left(1-\frac{\lambda^2}{\norm{\mathcal{N}(z_{ft})}^2},0 \right)$.
\end{itemize}
Simple LASSO shrinkage is identical to soft thresholding and corresponds to the proximal operator of $R=\norm{\cdot}_1$.
The Empirical Wiener, also known as the non-negative garrote
\cite{Gao1998:Wavelet.shrinkage.denoising.garrot},
is better than LASSO in terms of bias \cite{Rajmic2003:Exact.risk.analysis} but it still operates on coefficients individually.
EW is the proximal operator of a~function $R$ that has no explicit form, see for instance
\cite[Sec.\,3.2]{Antoniadis2007:Wavelet.methods.statistics} for more details.
In contrast to LASSO and EW, both WGL and PEW involve TF neighborhoods, such that the resulting value of the processed coefficient $z_{ft}$ depends on the energy contained in $\mathcal{N}(z_{ft})$.
Again, the difference is only the second power used by the PEW. 
It is interesting to note that the study \cite{Gribonval2018:Characterization.of.prox}
proves that neither WGL nor PEW are proximal operators of \emph{any} penalty $R$;
in other words, the respective shrinkages are purely heuristic.
Hand in hand with this, there are guarantees of convergence of Alg.\,\ref{alg:social.sparsity} for LASSO and the Empirical Wiener if the extrapolation parameter $\gamma$ is set properly
\cite{SiedenburgKowalskiDoerfler2014:Audio.declip.social.sparsity,combettes2011proximal},
while setting it in the case of WGL and PEW requires trial tuning.

The experiments in \cite{SiedenburgKowalskiDoerfler2014:Audio.declip.social.sparsity}
give evidence that out of the four choices, only PEW and EW are well-performing
and they outperform both the OMP \cite{Adler2011:Declipping} and C-IHT \cite{Kitic2013:Consistent.iter.hard.thresholding} approaches.


\subsection{Perceptual Compressed Sensing}
\label{sec:PerceptualCS.Defraene}

The approach by Defraene {\it et al.}
\cite{Defraene2013:Declipping.perceptual.compressed.sensing}
was for a~long time the only one
to include psychoacoustics in declipping
(both in the model itself and in the evaluation).
Although the approach is not quite related to the compressed sensing,
we refer to the method
using the same name as the authors coined.
The following optimization problem is solved:
\begin{equation}
    \label{eq:Perceptual.l1.Defraene}
    \min_{\z}
    \left\{
        \frac{1}{2} \norm{\Mr D\z - \Mr\y }_2^2 
        + \lambda \norm{\w\odot \z}_1
		\right\}\ 
		\text{s.t.}\  D\z\!\in\!\Gammah\!\cap\!\Gammal
    .
\end{equation}
This might look like just another variation of the synthesis-based declipping,
but the main difference from the other methods is that the weights $\w$ are computed based on a~human perception model.
Note that with respect to our terminology, this method is \emph{consistent in the clipped part}.

To be more specific about the method, the signal is processed window-by-window.
Ignoring the introduction of correct notation, task \eqref{eq:Perceptual.l1.Defraene} is solved independently for the signal chunks given by windowing.
The recovered signal is obtained by
\edt{the application of the synthesis $D$ to the found coefficients,}
and by reusing the reliable samples at positions given by the set $R$.
Once all the windows are processed this way, the final signal is obtained via the overlap--add procedure.


The psychoacoustic model enters the task through $\w$.
The authors of \cite{Defraene2013:Declipping.perceptual.compressed.sensing}
rely on the MPEG-1 Layer 1 psychoacoustic model 1 \cite{AudioSignalProcessingAndCoding,Shlien1994:Guide.to.MPEG-1}, 
which computes the instantaneous masking curve based on the incoming signal and on the absolute threshold of hearing.
In short, such a~curve informs us about the spectral components in the signal that will not be perceived when other strong components are present.
This effect is commonly known as the instantaneous masking or frequency masking.
Inspired by this effect, $\w$ is set as the inverse of this curve
(the curve in dB is non-negative, which justifies such an approach from the mathematical point of view).
Application of such weights can be interpreted as
discouraging the introduction of distinctively audible new spectral components
that are not present in the original signal.
On the other hand, the introduction of less audible or inaudible
spectral components
is tolerated to a~greater extent
\cite{Defraene2013:Declipping.perceptual.compressed.sensing}.

Worth noticing is that a~correctly treated masking curve:
1)~should be computed based on the original signal, not available in practice,
2)~should be applied to the very current window of the signal.
The authors of \cite{Defraene2013:Declipping.perceptual.compressed.sensing}
cope with both the issues in such a~way that they recurrently use the just declipped window as the base for calculating the masking curve,
which is then applied in declipping \edt{the subsequent window}.

In terms of the numerical treatment of \eqref{eq:Perceptual.l1.Defraene},
\cite{Defraene2013:Declipping.perceptual.compressed.sensing} proposes an algorithm termed PCSL1.
Its core, optimization part, refers to the CVX toolbox \cite{CVX}, 
but no particular codes for PCSL1 are available, unfortunately.
After several unsuccessful trials with CVX, we decided to solve 
the problem with the so-called generic proximal algorithm
introduced by Condat and V\~{u}
\cite{Condat2014:Generic.proximal.algorithm}, \cite{Vu2013:Splitting.algorithm}.
Such an algorithm, in the following abbreviated as the CV algorithm,
is able to solve convex problems with more than two terms, possibly even containing linear operators.
This is the case of \eqref{eq:Perceptual.l1.Defraene}, indeed.
Alg.\,\ref{alg:CV.for.perceptual.declipping} presents the particular shape of the CV algorithm for declipping.
The projection onto $\Gammah\cap\Gammal$ is done using the second and third lines of  \eqref{eq:proj_time_short}.
Algorithm \ref{alg:CV.for.perceptual.declipping} is guaranteed to converge if 
$\sigma<\tau^{-1}-1/2$.

\begin{algorithm}
    \setstretch{1.1}
    \DontPrintSemicolon
    \SetAlgoVlined
    \SetKwInput{Input}{Input}
    \SetKwInput{Par}{Parameters}
    \SetKwInput{Init}{Initialization}
    \Input{$D,\ \y\in\RR^N$,\ $\w\in\RR^P,\ \lambda>0,\ R,\ H,\ L$}
    \Par{$\sigma,\tau>0$ and $\rho\in(0,1]$}
    \Init{$\z^{(0)} \in \CC^P,\ \u^{(0)} \in \RR^N$}
    \setstretch{1.2}
    \For{$i=0,1,\dots$\,\upshape\textbf{until} convergence}{
			$\tilde{\z}^{(i+1)} =
			    \soft_{\tau\lambda\w}\!\left( \z^{(i)} \hspace{-.1em}-\hspace{-.1em} \tau D^* \! \left[ \Mr^* \Mr (D\z^{(i)} \hspace{-.1em}-\hspace{-.1em} \y) \hspace{-.1em}+\hspace{-.1em} \u^{(i)} \right] \right)$\; 
			$\z^{(i+1)} = \rho\tilde{\z}^{(i+1)} + (1-\rho)\z^{(i)}$\;
			$\p^{(i+1)} = \u^{(i)} + \sigma D(2\tilde{\z}^{(i+1)} - \z^{(i)})$ \comment{auxiliary}\;
			$\tilde{\u}^{(i+1)} = \p^{(i+1)} - \sigma\,\proj_{\Gammah\cap\Gammal}\left( \p^{(i+1)}/\sigma \right)$\;
			$\u^{(i+1)} = \rho\tilde{\u}^{(i+1)} + (1-\rho)\u^{(i)}$\;		
    }
    \KwRet{$D\z^{(i+1)}$}
    \caption{Condat--V\~{u} (CV) algorithm solving \eqref{eq:Perceptual.l1.Defraene}
    }
    \label{alg:CV.for.perceptual.declipping}	
\end{algorithm}

\subsection[Psychoacoustically motivated l1 minimization]{Psychoacoustically motivated $\ell_1$ minimization}
%
The second method that involves psychoacoustics, by Záviška {\it et al.} \cite{ZaviskaRajmicSchimmel2019:Psychoacoustics.l1.declipping}, is similar to the above (Section \ref{sec:PerceptualCS.Defraene}),
but it is designed as completely consistent.
Recall that this means that the declipped signal should belong to the set $\Gamma$ defined in \eqref{eq:gamma.subsets}.
%
The problem solved in \cite{ZaviskaRajmicSchimmel2019:Psychoacoustics.l1.declipping}
is actually identical to 
\eqref{eq:problem_syn.basic},
but the weights are now derived from the human perception model.
It is a~synthesis-based signal model, and again,
the Douglas--Rachford algorithm presented in Alg.\,\ref{alg:DR.declipping}
can be used to find the numerical solution
(with the efficient projection onto $\Gamma^*$ in the case that $D$ is a~tight frame).

Unlike
\cite{Defraene2013:Declipping.perceptual.compressed.sensing},
the paper \cite{ZaviskaRajmicSchimmel2019:Psychoacoustics.l1.declipping} discusses multiple ways of choosing the weights $\w$. 
Besides the basic inversion, there are several other options of \qm{inverting} the masking curve that have been introduced and evaluated.
Surprisingly, the best declipping results were obtained using weights which simply grow quadratically with frequency!
Such an option is not psychoacoustically inspired at all,
but its success might be explained by the fact that signals after clipping have a~very rich spectrum,
while the spectra of original signals decay with increasing frequency.
In other words, regularizing the spectrum in such a~way (in addition to sparsity)
seems to be more powerful than modeling delicate perceptual effects.
For the experiments in Sec.\,\ref{sec:Evaluation}, just this \qm{parabola} option was selected.

Motivated by such an interesting observation, we also employed these quadratic weights $\w$ in the method from Sec.\,\ref{sec:PerceptualCS.Defraene}.
See more in the evaluation part of the article.

\subsection{Dictionary Learning approach}
\label{ssec::dict_learn}
In the next two sections, we consider $T$ windows $\y_1, \dots, \y_T$ of the clipped signal $\y$,
and their corresponding
consistency sets
$\Gamma_t=\Gamma(\y_t)$.

Sparsity-based methods reviewed so far use fixed and known synthesis operators, such as the DCT or Gabor transforms.
However, another approach consists in \emph{adapting}\/ $D$ to the observed data.
Learning the dictionary
(we prefer the term \emph{dictionary} since $D$ will be treated in its matrix form from now on),
rather than using an off-the-shelf one, has shown improvements in inverse problems such as denoising or inpainting \cite{Mairal2014.Sparse, eladbook}.
Dictionary learning (DL) from clipped observations has been formulated by Rencker {\it et al.} \cite{RenckerBachWangPlumbley2018:Sparse.recovery.dictionary.learning, RenckerBachWangPlumbley2018:Consistent.dictionary.learning.LVA}.
Given a~collection of $T$ clipped signals $\y_1, \dots, \y_T$
(typically corresponding to $T$ overlapping time windows extracted from a signal),
dictionary learning from clipped observations can be formulated as:
\begin{equation}\label{eq:DL_problem}
    \min_{\z_t, D}\, \sum_{i = 1}^T \dEU(D\z_t, \Gamma_t)^2 \ \text{s.t.} \ \|\z_t\|_0 \leq K,\ t=1,\ldots,T,
\end{equation}
where $\dEU(D\z_t, \Gamma_t)$ is the Euclidean distance of $D\z_t$ to the set $\Gamma_t$, and $\Gamma_t$ is the feasibility set corresponding to the signal $\y_t$ (as defined in \eqref{eq:gamma.subsets}).
Note that using the notation in Sec.\,\ref{sec:problem_formulation}, $\dEU(\cdot, \Gamma_t)^2$ is equivalent to the data-fidelity term in \eqref{eq:Social.sparsity.problem}, and is convex, differentiable with Lipschitz gradient thanks to the convexity of $\Gamma_t$.
DL
algorithms typically alternate between optimizing $\z_1, \dots, \z_T$ with $D$ fixed (sparse coding step), and optimizing $D$ with $\z_1, \dots, \z_T$ fixed (dictionary update step)~\cite{eladbook}.

The sparse coding step solves, for each $t$ independently: 
\begin{equation}\label{eq:DL_sparse_coding_step}
\min_{\z_t} \dEU(D\z_t, \Gamma_t)^2 \quad \text{s.t.} \quad \|\z_t\|_0 \leq K,
\end{equation} 
which can be aproximated using consistent Iterative Hard Thresholding (IHT).
Consistent IHT, proposed in \cite{Kitic2013:Consistent.iter.hard.thresholding}, is a~simple algorithm that iterates:
\begin{equation}\label{eq:consistentIHT}
    \z_t \gets \mathcal{H}_K(\z_t + \mu D^\top\! (D\z - \proj_{\Gamma_t}(D\z_t)),
\end{equation}
%
which corresponds to a~gradient descent step (with the parameter $\mu$), followed by the hard thresholding.
The $\ell_0$ constraint in \eqref{eq:DL_problem} can also be relaxed into an $\ell_1$ constraint,
in which case the sparse coding step would correspond to an ISTA-type algorithm
in Alg.\,\ref{alg:social.sparsity}.

The dictionary update step is formulated as:
\begin{equation}
    \min_{D \in \mathcal{D}} \sum_{t=1}^T \dEU(D\z_t, \Gamma_t)^2,
\end{equation}
which can be solved using (accelerated) gradient descent.
Note that $D$ is constrained to belong to
$\mathcal{D} = \{ D = \left[\d_1, \dots, \d_P\right] \ | \ \|\d_p\|_2 \leq 1 \ \text{for} \ p = 1, \dots, P\}$
in order to avoid scaling ambiguity.

The overall dictionary learning algorithm is presented in Algorithm \ref{alg:DL_algorithm}. When $\y_1, \dots, \y_T$ correspond to overlapping windows extracted from a given signal, each window can be recovered using the estimated dictionary and sparse coefficients as $\hat{D}\hat{\z}_1, \dots, \hat{D}\hat{\z}_T$. The overall signal can then be estimated using overlap--add.

\begin{algorithm}
    \setstretch{1.1}
	\DontPrintSemicolon
	\SetAlgoVlined
	\SetKwInput{Input}{Input}
    \SetKwInput{Par}{Parameters}
    \SetKwInput{Init}{Initialization}
	\Input{$\y_1, \dots, \y_T\in\RR^N,\ R,\ H,\ L$}
	\Par{$K$, $P$}
	\Init{$\z_1^{(0)}, \dots, \z_T^{(0)} \in \RR^P,\ D^{(0)} \in \RR^{N \times P}$}
	\setstretch{1.2}
    \For{$i=0,1,\dots$\,\upshape\textbf{until} convergence}{	
        Solve for $t = 1, \dots, T$, using e.g., consistent IHT:
		$\z_t^{(i+1)} = \argmin_{\z_t} \dEU(D^{(i)}\z_t, \Gamma_t)^2 \ \text{s.t.} \ \|\z_t\|_0 \leq K$  \;
		Solve using, e.g., accelerated gradient descent:
		$D^{(i+1)} = \argmin_{D} \sum_{t=1}^T \dEU(D\z_t^{(i+1)}, \Gamma_t)^2$  \;
		
	}
	\KwRet{$D^{(i+1)}\z_1^{(i+1)}, \dots, D^{(i+1)}\z_T^{(i+1)}$}
	\caption{Dictionary learning algorithm for declipping \cite{RenckerBachWangPlumbley2018:Consistent.dictionary.learning.LVA}}	\label{alg:DL_algorithm}
\end{algorithm}


\subsection{Nonnegative Matrix Factorization}
\label{sec::nmf_declipping}

Another approach proposed recently by Bilen {\it et al.} \cite{BilenOzerovPerez2015:declipping.via.NMF,Bilen2018:NTF_audio_inverse_problems} is based on the nonnegative matrix factorization (NMF).
This is a~dictionary learning approach too, but instead of learning a~dictionary of the waveform,
it learns a nonnegative dictionary together with nonnegative decomposition coefficients to approximate the unknown power spectrogram of the original signal.
This corresponds to
the assumption that the power spectrogram is approximately low-rank.
Given that the power spectrogram is a phase-free representation, this modeling is phase-invariant, thus allowing using a~dictionary of a considerably smaller size than the dictionary size in the approach presented in Section~\ref{ssec::dict_learn}.

NMF modeling is defined on the latent clean signal power spectrogram 
obtained
from the analysis short-time Fourier transform (STFT) coefficients.
Note that it is also possible to decompose power spectrograms of synthesis coefficients, as in \cite{fevotte2018estimation},
though this has not yet been done for audio declipping but for a~related problem of compressed sensing recovery \cite{fevotte2018estimation}.
More specifically, the analysis NMF approach assumes that the power spectrogram nonnegative matrix ${\bf P} = [p_{ft}]_{f,t=1}^{F,T}$
(with $p_{ft}=|z_{ft}|^2$, and $z_{ft}$ being clean signal STFT coefficients)
is approximated as
\begin{equation}
    {\bf P} \approx {\bf V} = {\bf W} {\bf H},
    \label{eq::nmf_decomp}
\end{equation}
with ${\bf W} \in \RR_+^{F \times K}$ and ${\bf H} \in \RR_+^{K \times T}$ nonnegative matrices, and $K > 0$ usually chosen much smaller than $F$ and $T$, $K \ll \min(F, T)$.
The matrix ${\bf W}$
can be understood as the power spectrum dictionary
(columns of ${\bf W}$ being characteristic spectral patterns),
while ${\bf H}$ contains the corresponding nonnegative decomposition (activation) coefficients.

Note that the modeling in \eqref{eq::nmf_decomp} is not yet well defined, since the approximation is not specified mathematically and the power spectrogram ${\bf P}$ is unknown.
To specify it properly, it is proposed in \cite{BilenOzerovPerez2015:declipping.via.NMF,Bilen2018:NTF_audio_inverse_problems} to resort to the maximum likelihood (ML) optimization under a probabilistic Gaussian formulation of Itakura Saito (IS) NMF \cite{fevotte2009nonnegative}.
To simplify the formulation here, let all signals be considered either in the time domain (windowed, with overlap) or in the frequency
domain, while the two domains are related by the DFT for each time block separately.
The DFT, denoted $A$ here, $A\colon\CC^F \to \CC^F$, is unitary,
and the number of frequency channels $F$ is identical to the number of time-domain samples.
Let ${\bf Y} = [\y_1, \dots, \y_T]$ and ${\bf X} = [\x_1, \dots, \x_T]$
denote the windowed versions of the clipped and original (unknown) signals, respectively,
and ${\bf Z} = [\z_1, \dots, \z_T]$ the STFT of the original signal ($\z_t = A \x_t$).
It is assumed that the
coefficients in $\bf Z$ are all mutually independent, and each coefficient $z_{ft}$ follows a complex circular zero-mean Gaussian distribution
$
    z_{ft} \sim {\mathcal N}_{\mathrm{c}} (0, v_{ft})
$
%
with ${\bf V} = [v_{ft}]_{f,t}$ being a low-rank power spectrogram approximation specified in \eqref{eq::nmf_decomp}. NMF model parameters are estimated optimizing the ML criterion (see \cite{BilenOzerovPerez2015:declipping.via.NMF} for details)
\begin{equation}
    ({\bf W}, {\bf H}) = {\rm arg} \max_{{\bf W}', {\bf H}'} \ p({\bf Y} | {\bf W}', {\bf H}')
\end{equation}
via the generalized expectation--maximization (GEM) algorithm \cite{dempster1977maximum} with multiplicative update (MU) rules \cite{fevotte2009nonnegative}.
The final windowed signal block estimate $\widehat{\bf X}$ is recovered via Wiener filtering, see \eqref{eq::nmf_Wiener}.%
\footnote{Estimating windowed signal blocks results in a problem relaxation \cite{Bilen2018:NTF_audio_inverse_problems} since the overlapping frames are clearly not independent, but those dependencies are not exploited during estimation.}
This is altogether summarized in Alg.\,\ref{alg:NMF_algorithm},
where $\Mrt$ denotes the restriction of the operator $\Mr$ to block $t$, 
and all operators (e.g., $\Mrt$ or $A^*$), when applied to matrices, are applied column-wise.
It should be highlighted that, though the NMF modeling \eqref{eq::nmf_decomp} is defined on the signal power spectrogram,
the signal is reconstructed with both amplitude and phase since the Wiener filtering \eqref{eq::nmf_Wiener}
with a~complex-valued Wiener gain
(matrix $\boldsymbol{\Sigma}^*_{\Mrt \y_t \z_t} \boldsymbol{\Sigma}^{-1}_{\Mrt \y_t \Mrt \y_t}$)
maps from the time domain to the complex-valued STFT domain.

\begin{algorithm}
    \setstretch{1.1}
	\DontPrintSemicolon
	\SetAlgoVlined
	\SetKwInput{Input}{Input}
    \SetKwInput{Par}{Parameters}
    \SetKwInput{Init}{Initialization}
	\Input{$\y_1, \dots, \y_T\in\RR^F,\ R,\ H,\ L$}
	\Par{$K > 0$}
	\Init{${\bf W}^{(0)} \in \RR_+^{F \times K},\ {\bf H}^{(0)} \in \RR_+^{K \times T}$ (random)}
	\setstretch{1.2}
	${\bf V}^{(0)} = {\bf W}^{(0)} {\bf H}^{(0)}$ \;
    \For{$i=0,1,\dots$\,\upshape\textbf{until} convergence}{
        Estimate posterior power spectrogram ${\bf P} = [p_{ft}]$: \label{alg::step_1}
        \newline
        $\hat{p}_{ft} = |\hat{z}^{(i+1)}_{ft}|^2 + \widehat{\boldsymbol{\Sigma}}_{\z_t \z_t}(f,f)$,
        \newline
        where $(f,f)$ picks the $f$-th diagonal matrix entry
        and%
        \begin{equation}
          \hat{\z}^{(i+1)}_t = \boldsymbol{\Sigma}^*_{\Mrt \y_t \z_t} \boldsymbol{\Sigma}^{-1}_{\Mrt \y_t \Mrt \y_t} \Mrt \y_t,
          \label{eq::nmf_Wiener}
        \end{equation}
        $\widehat{\boldsymbol{\Sigma}}_{\z_t \z_t} = \boldsymbol{\Sigma}_{\z_t \z_t} -
        \boldsymbol{\Sigma}^*_{\Mrt \y_t \z_t} \boldsymbol{\Sigma}^{-1}_{\Mrt \y_t \Mrt \y_t}
				\boldsymbol{\Sigma}_{\Mrt \y_t \z_t}$,  \newline
        with \newline
        $ \boldsymbol{\Sigma}_{\z_t \z_t} = {\mathrm{diag}} \left( [v^{(i)}_{ft}]_f \right), \,$
        $ \boldsymbol{\Sigma}_{\Mrt \y_t \z_t} = \Mrt A^* \, \boldsymbol{\Sigma}_{\z_t \z_t} $, \newline
        $\boldsymbol{\Sigma}_{\Mrt \y_t \Mrt \y_t} = \Mrt A^* \boldsymbol{\Sigma}^*_{\Mrt \y_t \z_t}$. 
        
        Compute: $\hat{\x}^{(i+1)}_1 = A^* \hat{\z}^{(i+1)}_1, \dots, \hat{\x}^{(i+1)}_T = A^* \hat{\z}^{(i+1)}_T$ \label{alg::step_2}
        
        Update NMF parameters using MU rules:
        $ {\bf W}^{(i+1)} = {\bf W}^{(i)} \odot \frac{\left( \left[{\bf W}^{(i)} {\bf H}^{(i)}\right]^{.-2} \odot \widehat{\bf P} \right) \left( {\bf H}^{(i)} \right)^\top}
        {\left[{\bf W}^{(i)} {\bf H}^{(i)}\right]^{.-1} \left( {\bf H}^{(i)} \right)^\top}, $
        $ {\bf H}^{(i+1)} = {\bf H}^{(i)} \odot \frac{\left( {\bf W}^{(i+1)} \right)^\top \left( \left[{\bf W}^{(i+1)} {\bf H}^{(i)}\right]^{.-2} \odot \widehat{\bf P} \right)}
        {\left( {\bf W}^{(i+1)} \right)^\top \left[{\bf W}^{(i+1)} {\bf H}^{(i)}\right]^{.-1}}, $ \newline
        with $\odot$ and $\left[\cdot\right]^{.b}$ denoting element-wise matrix product and power, all divisions being element-wise as well,
        and $\left[\cdot\right]^\top$ denoting the transpose.
        
        Update: ${\bf V}^{(i+1)} = {\bf W}^{(i+1)} {\bf H}^{(i+1)}$
	}
	\KwRet{$\hat{\x}^{(i+1)}_1, \dots, \hat{\x}^{(i+1)}_T$ {\rm after applying the corresponding synthesis window and overlap--add to get the signal in time domain.}}
	\caption{NMF GEM algorithm \cite{BilenOzerovPerez2015:declipping.via.NMF}}
	\label{alg:NMF_algorithm}
\end{algorithm}

Note that the consistency in the clipped part
is not satisfied in Algorithm~\ref{alg:NMF_algorithm},
and it is difficult to satisfy it properly since with this constraint the posterior distribution of ${\x}_t$
(given the observations and the NMF model)
is not Gaussian any more.
To take the constraint into account, an ad hoc strategy has been proposed in \cite{BilenOzerovPerez2015:declipping.via.NMF,Bilen2018:NTF_audio_inverse_problems}.
This strategy consists in checking in step~\ref{alg::step_2} whether the clipping constraint is satisfied.
If not, the samples of those blocks ${\hat\x}_t$ for which it is not satisfied are projected on the corresponding clipping thresholds, dynamically added to the reliable set,
and for those blocks the steps~\ref{alg::step_1} and \ref{alg::step_2} are repeated again, and, if necessary, iterated till clipping constraint is completely satisfied in step~\ref{alg::step_2}.
Whenever the iteration of the main algorithm starts over again, the reliable set is re-initialized (see \cite{BilenOzerovPerez2015:declipping.via.NMF,Bilen2018:NTF_audio_inverse_problems}).


\subsection{Janssen's autoregressive interpolation}
The Janssen method \cite{javevr86} published back in 1986 and thoroughly discussed recently in \cite{Oudre2018:Janssen.implementation}
relies on the autoregressive (AR) signal model.
It assumes that a particular signal sample can be induced via a~fixed linear combination of preceding samples.
The coefficients in such a combination are the AR coefficients
and their total number is called the order of the AR model.
The model can be alternatively interpreted such that the audio signal is generated by a Gaussian white noise filtered by an all-pole filter.

In practice, the AR model can be successfully applied to signals containing harmonic components.
The Janssen method cannot handle the clipping constraints;
hence in declipping, it is only possible to use it in order to replace the clipped samples by the values linearly estimated from the reliable samples.
In that regard, the Janssen method 
\edt{actually belongs to the audio inpainting methods.}

Despite its simplicity and age, the algorithm is a~strong competitor of the most recent audio inpainting methods,
\cite{MokryZaviskaRajmicVesely2019:SPAIN}.
That is why we decided to consider it within our evaluation.

\section{Evaluation}
\label{sec:Evaluation}
This section compares the selected audio declipping methods \edt{from the previous section} in terms of the quality of reconstruction. 
First, the design of the experiments
is described, 
along with the characterization of the audio dataset. 
The evaluation metrics used to objectively assess the quality of reconstruction are presented then.
Subsection \ref{subsec:algorithms.settings} contains details  about the algorithms from the practical viewpoint, such as settings of the parameters and comments on the behavior of the algorithms.
Finally,
the results are presented and discussed.

\subsection{Experiment design and the dataset}

The audio database used for the evaluation consists of 10~musical excerpts in mono, sampled at 44.1\,kHz, with an approximate duration of 7~seconds.
They were extracted from the EBU SQAM database.\footnote{https://tech.ebu.ch/publications/sqamcd} 
The excerpts were thoroughly selected to cover a~wide range of audio signal characteristics.
Since a~significant number of the methods are based on signal sparsity, the selection took care that different levels of sparsity were  included in the signals
(w.r.t.\ the Gabor transform).
\edt{We have selected the sounds of the violin, clarinet, basson, harp, glockenspiel, celesta, accordion, guitar, piano, and the wind ensemble.}

To the best of our knowledge,
the only declipping experiments including audio sampled at 44.1\,kHz were carried in
\cite{Defraene2013:Declipping.perceptual.compressed.sensing} and
\cite{ZaviskaRajmicSchimmel2019:Psychoacoustics.l1.declipping},
while the others used audio at 16\,kHz at the most.
This survey thus provides the very first large-scale experiment for a~high-quality sampled audio.

The input data were clipped in agreement with the model in Eq.\,\eqref{eq:clipping}, using clipping levels
that were chosen to lead to 7 different input signal-to-distortion ratios (SDR).
The SDR for two signals $\u$ and $\vv$ is defined as
\begin{equation}
    \sdr (\u,\vv) = 20\log_{10} \frac{\norm{\u}_2}{\norm{\u-\vv}_2}.
    \label{eq:SDR}
\end{equation}
Recall that $\x$ denotes the original and $\y$ the clipped signal;
hence, the input SDR is computed as $\sdr (\x,\y)$.

\begin{figure}
    \centering
%
%
%
\definecolor{c1}{rgb}{0,	0,	0.531250000000000}
\definecolor{c2}{rgb}{0,	0.281250000000000,	1}
\definecolor{c3}{rgb}{0,	0.750000000000000,	1}
\definecolor{c4}{rgb}{0.218750000000000,	1,	0.781250000000000}
\definecolor{c5}{rgb}{0.687500000000000,	1,	0.312500000000000}
\definecolor{c6}{rgb}{1,	1,	0}
\definecolor{c7}{rgb}{1,	0.8500,	0}
\definecolor{c8}{rgb}{1,	0.5625,	0}
\definecolor{c9}{rgb}{0.937500000000000,	0,	0}
\definecolor{c10}{rgb}{0.500000000000000,	0,	0}
\begin{tikzpicture}[scale=0.57]

\begin{axis}[%
width=5.356in,
height=3.3in,
scale only axis,
xmin=0,
xmax=21,
xtick={ 1,  3,  5,  7, 10, 15, 20},
xlabel style={font=\Large\color{white!15!black}},
xticklabel style={font=\large},
xlabel={input SDR (dB)},
ymin=0,
ymax=90,
ylabel style={font=\Large\color{white!15!black}},
ylabel={clipped samples (\%)},
yticklabel style={font=\large},
axis background/.style={fill=white},
xmajorgrids,
ymajorgrids,
legend style={legend cell align=left, align=left, draw=white!15!black},every axis plot/.append style={very thick}
]
\addplot [color=c1, mark=o, mark options={solid, c1}]
  table[row sep=crcr]{%
1	70.7811572146633\\
3	48.1171167955453\\
5	34.3747464881511\\
7	25.5166774290897\\
10	16.3389806308835\\
15	7.19178637834189\\
20	2.96166738596429\\
};
\addlegendentry{violin}

\addplot [color=c2, mark=o, mark options={solid, c2}]
  table[row sep=crcr]{%
1	83.3001600525301\\
3	63.5155469829414\\
5	48.6751207233827\\
7	38.1215715243293\\
10	26.3638664313758\\
15	12.4916211816528\\
20	5.1004774216495\\
};
\addlegendentry{clarinet}

\addplot [color=c3, mark=o, mark options={solid, c3}]
  table[row sep=crcr]{%
1	80.4189061300287\\
3	55.6705890755852\\
5	38.3915147402818\\
7	26.9999821438137\\
10	16.2194882417013\\
15	7.24639751441887\\
20	3.31410817277646\\
};
\addlegendentry{bassoon}

\addplot [color=c4, mark=o, mark options={solid, c4}]
  table[row sep=crcr]{%
1	86.2905452852668\\
3	56.9784047782486\\
5	34.8545458849149\\
7	22.8241058536701\\
10	12.289235527572\\
15	4.61847389558233\\
20	1.81788055956636\\
};
\addlegendentry{harp}

\addplot [color=c5, mark=o, mark options={solid, c5}]
  table[row sep=crcr]{%
1	77.8522623338893\\
3	45.9826760639156\\
5	28.7757572496906\\
7	19.1811481142734\\
10	9.61586054769463\\
15	2.9663205466186\\
20	0.552805724431054\\
};
\addlegendentry{glockenspiel}

\addplot [color=c6, mark=o, mark options={solid, c6}]
  table[row sep=crcr]{%
1	83.1702947793821\\
3	58.793765781226\\
5	42.1513884203873\\
7	29.6200715906821\\
10	17.0055707113546\\
15	6.15267220572005\\
20	2.15689035752767\\
};
\addlegendentry{celesta}

\addplot [color=c7, mark=o, mark options={solid, c7}]
  table[row sep=crcr]{%
1	82.0418891026353\\
3	58.7178332738871\\
5	42.8345894996579\\
7	30.7938113587061\\
10	18.5507212480193\\
15	7.96881121252697\\
20	3.29195488326132\\
};
\addlegendentry{accordion}

\addplot [color=c8, mark=o, mark options={solid, c8}]
  table[row sep=crcr]{%
1	83.2820407954604\\
3	59.9291106453317\\
5	42.086464563843\\
7	27.8953018761822\\
10	14.4934648876165\\
15	4.50590460610398\\
20	1.70781998193685\\
};
\addlegendentry{guitar}

\addplot [color=c9, mark=o, mark options={solid, c9}]
  table[row sep=crcr]{%
1	80.2671254277357\\
3	55.3901288571504\\
5	38.3917513354389\\
7	26.367342636546\\
10	15.0977063201364\\
15	5.71518918247422\\
20	1.93153969801576\\
};
\addlegendentry{piano}

\addplot [color=c10, mark=o, mark options={solid, c10}]
  table[row sep=crcr]{%
1	81.2716716716717\\
3	54.2574574574575\\
5	36.7827827827828\\
7	24.8236236236236\\
10	13.5127127127127\\
15	4.75475475475475\\
20	1.66966966966967\\
};
\addlegendentry{wind ensemble}

\end{axis}
\end{tikzpicture}%
    \caption{Percentages of the clipped samples for the selected input SDRs.}
    \label{fig:perc_clip}
\end{figure}
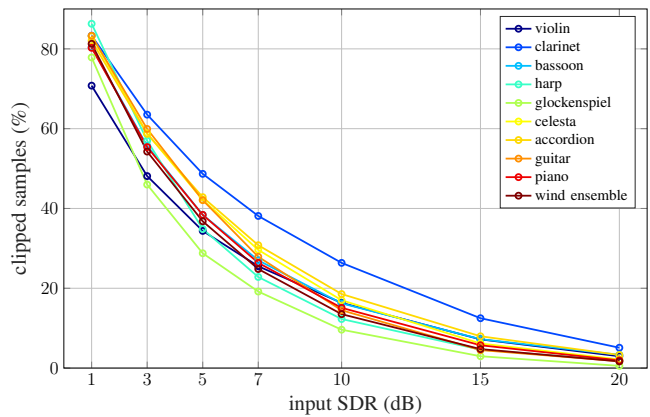

With respect to the human perception of clipping severity,
the SDR is more meaningful than treating signals according to the clipping levels or according to the percentage of clipped samples \cite{Gaultier2019:PhD.Thesis}.
The particular input SDR levels are chosen
to cover the range from very harsh clipping to mild but still noticeable clipping.
The specific values along with the respective percentages of clipped samples
are visualized in Fig.\,\ref{fig:perc_clip}.
%
Since the input SDR is used,
there is no need to peak-normalize the audio samples before processing because the number of the clipped samples remains the same, independently of scaling. 


All the data have been processed and evaluated using MATLAB in double precision, and therefore there is no additional distortion caused by quantization during the process.

\subsection{Evaluation metrics}
\label{subsec:evaluation.metrics}
To evaluate the reconstruction quality, we use several metrics.

\subsubsection{Signal-to-Distortion Ratio}
Firstly, we utilize the signal-to-distortion ratio (SDR),
which is one of the simplest, nonetheless, one of the most used methods.
It expresses the physical quality of reconstruction, i.e.,
how (numerically) close to the ground truth $\x$ the recovered signal $\hat{\x}$ is.

The restored signal $\hat\x$ is evaluated 
using the output SDR, which is
computed as $\sdr (\x,\hat\y)$ using \eqref{eq:SDR}.
%
%
Note that evaluating the SDR on the whole signal
may handicap the methods that produce signals inconsistent in the reliable part. 
(Some of such methods may include replacing the reliable samples with the reliable samples from the input clipped signal $\y$ as the final step of the reconstruction.)
Therefore, as for example in \cite{SiedenburgKowalskiDoerfler2014:Audio.declip.social.sparsity},
we compute the SDR on the clipped part only, $\sdr_\mathrm{c}$, as
%
\begin{equation}
    \sdrc (\x,\hat\x) = 20\log_{10} \frac{\left\|
    \left[{\Mh}\atop{\Ml}\right]\x\right\|_2}
    {\left\|\left[{\Mh}\atop{\Ml}\right]\x-\left[{\Mh}\atop{\Ml}\right]\hat\x\right\|_2}.
    \label{eq:SDRc}
\end{equation}
Since this article aims at signal reconstruction, we will rather 
use the SDR improvement,
i.e., the difference between the SDR of the restored and the clipped signal,
formally defined as
\begin{equation}
    \dsdrc =
    \sdrc (\x, \hat{\x}) - \sdrc (\x, \y),
    \label{eq:dSDR}
\end{equation}
and similarly for $\dsdr$.
%
%
%
Note that this criterion does not take into consideration whether a~method is or is not consistent in the clipped part,
since \eqref{eq:SDRc} does not either.
Note also that in the case of consistency in the reliable part,
the \dsdr{} produces the same values, no matter whether the $\sdr$ is computed
on the whole signal or on the clipped samples only,
and $\dsdr$ and $\dsdrc$ simply coincide.

\subsubsection{PEAQ}
A good quality assessment should correspond as much as possible to
perceptual experience.
From this point of view, the SDR is not the best metric, since the physical similarity with the original signal does not automatically imply perceptual similarity and quality. 
Hence, an evaluation metric involving the human perceptual system should be used. 

PEAQ---Perceptual Evaluation of Audio Quality, which became the ITU-R recommendation (BS 1387) in 1999, is considered standard for audio quality evaluation.
The final output of PEAQ is the Objective Difference Grade (ODG) rating the perceived difference between the clean and the degraded signal.
The ODG score is shown in Table~\ref{tab:ODG}.

For the experiments, we used the MATLAB implementation,\!%
\footnote{\url{http://www-mmsp.ece.mcgill.ca/Documents/Software/}}
implemented according to the revised version of PEAQ (BS.1387-1) and available with a detailed review of this method \cite{Kabal2002:PEAQ}.
Unfortunately, this implementation is suited only for signals with the 48\,kHz sampling frequency.
Since the audio database used is sampled at 44.1\,kHz, we upsample the signals in order to compute the PEAQ ODG.

\begin{table}[h!]
    \small
    \centering
    \caption{Objective difference grade}
    \begin{tabular}{|Sr|Sl|}\hline
    \rowcolor{white!90!black}
    \multicolumn{1}{|Sc|}{ODG} & \multicolumn{1}{|Sc|}{Impairment description}\\ \hline\hline
    $0.0$ & Imperceptible  \\
    \rowcolor{white!95!black}
    $-1.0$ & Perceptible, but not annoying \\ 
    $-2.0$ & Slightly annoying \\
    \rowcolor{white!95!black}
    $-3.0$ & Annoying \\ 
    $-4.0$ & Very annoying \\ \hline
    \end{tabular}
    \label{tab:ODG}
\end{table}

\subsubsection{PEMO-Q}
As another evaluation metric taking into account the human auditory system we use the PEMO-Q method published in \cite{Huber:2006a}. 
The Matlab implementation%
\footnote{\url{https://www.hoertech.de/de/f-e-produkte/pemo-q.html}}
is freely available for academic use and research. 
PEMO-Q computes the perceptual similarity measure (PSM), which can be mapped to the ODG score (see Table~\ref{tab:ODG}). 
The mapping to ODG is available only for signals with 44.1\,kHz sampling frequency but since this is the case of the database used, no additional resampling is required.

\subsubsection{\edt{\Rnonlin{}}}

\edt{
Even though PEMO-Q and PEAQ provide a~reasonable prediction of perceptual evaluation
and are often used in restoration studies, they were actually designed for the evaluation of bit-rate reduction systems in audio coding.
For this reason, we also include the \Rnonlin{} metric \cite{TanMoore2004:Rnonlin.paper}, designed specifically for evaluating nonlinear distortions.
}%

\edt{
The \Rnonlin{} metric can be used to predict the perceived quality of nonlinearly distorted signals.
The prediction, which is a~value on a scale from 0 to 1 (worst to best),
is based on the weighted cross-correlation of the output of an array of gammatone filters.
The Matlab implementation is available at Matlab central.\!%
\footnote{\url{https://www.mathworks.com/matlabcentral/fileexchange/50230-rnonlin_calc}}
%
}

\subsection{Algorithms and settings}
\label{subsec:algorithms.settings}
Parameter fine-tuning is a necessary part of the experiment,
as the overall results depend highly on the parameter selection. 
In our experiments, we attempted to tune the parameters of each algorithm 
such that it produces the best possible reconstruction result in terms of the SDR.

Several above-presented algorithms employ a~time-frequency (TF) transform and\,/\,or processing by blocks.
For such cases, we tried to unify the related setting across the algorithms to ensure a~fair comparison. 
The optimal way would be to tune the parameters for every input SDR separately
(for instance, harsh clipping with a~great number of clipped samples benefits from the use of longer windows compared to mild clipping). 
For simplicity, we stick to a~compromise among all the cases,
and the parameters of each method 
stay fixed for all test signals and all clipping levels. 
Specifically, if the algorithm processes the signal block-by-block, 8192-sample-long ($\thicksim$186\,ms) blocks are used. 
If the algorithm utilizes the DGT, we use the 8192-sample-long Hann window with 75\% overlap and 16384 frequency channels.
Unfortunately, such a~setting could not be used
in C-OMP, NMF and
DL
due to the high computational complexity of these algorithms.
For \mbox{C-OMP} and
DL, we use windows of 1024 samples, with 75\% overlap, and twice-redundant dictionaries of size $P = 2048$.
The NMF algorithm use windows of size 2048, with 2048 frequency channels.

The implementation of
the
TF transforms is handled by the LTFAT toolbox \cite{LTFAT} in most of the methods.

As for the termination criterion,
we resort to using just a~simple maximum number of iterations.
This number has been empirically set for each algorithm independently to make sure that the algorithm fully converged.
The only exception is the SPADE algorithms that have the  $\epsilon$ parameter involved straight in the problem formulation, see
\eqref{eq:SPADE.analysis.formulation} and
\eqref{eq:SPADE.synthesis.formulation},
and $\epsilon$ is thus naturally used as the termination threshold.


If an algorithm produces a~signal inconsistent in the reliable part, we do not replace that part with the original reliable samples before the evaluation.
Naturally, such a~replacement increases the overall output SDR,
\edt{as discussed in Sec.\,\ref{subsec:evaluation.metrics}}.
\edt{In terms of perceptual metrics,
evidence suggests that doing this replacement is not beneficial for signals with low input SDR.
This is probably due to the fact that the replacement of the samples creates} discontinuities in the waveform,
leading to the introduction of artificial higher harmonics that in the end degrade the reconstruction quality.
\edt{The advantage of matching the reliable samples is, however, pronounced in the high input SDR regime,
where there is only a~small number of clipped samples and the effect of new discontinuities is negligible.}
Note that these additional variants are excluded from the comparison for clarity;
\edt{however, the supplementary repository presents the described effect in the form of graphs.}




\subsubsection{Constrained OMP}
C-OMP was tested using the implementation provided by the authors of
\cite{Adler2011:Declipping, Adler2012:Audio.inpainting}
in the SMALLbox MATLAB toolbox.\footnote{\url{http://small.inria.fr/software-data/smallbox/}}
We have used the DGT-based implementation with min-constraint, as we have found that it provided a~reasonable
trade-off between performance and computational complexity.
Note that when the clipping level is low (many samples missing),
the constrained optimization (the final step of Alg.\,\ref{alg:Constrained.OMP}) often failed to converge.
We believe this is because the support set $\Omega$ estimated
with the OMP
(without any clipping constraint) is often suboptimal, leading to a~signal that is clipping-\emph{inconsistent}.
As a result, a signal that belongs to the range space of $D_{\Omega}$ and the clipping consistency set $\Gammah \cap \Gammal$ might not exist or have a very large amplitude,
thus making the constraint $D_{\Omega}\z_{\Omega} \in \Gammah \cap \Gammal$ unfeasible.
In that scenario, following the guidelines by the authors in \cite{Adler2012:Audio.inpainting} (implemented in the accompanying code), we simply return the output of the (unconstrained) OMP as a~solution.

\subsubsection{SPADE}
Although we have the original implementation of A-SPADE,
we use our own implementation, which is slightly improved compared to  the original version. 
The main differences are the means by which the signal is windowed
and the hard-thresholding step, where we take into account the complex conjugate structure of the DFT and always process pairs of  coefficients
(hence producing purely real signals with the inverse DFT). 
The above also applies to the S-SPADE implementation.

Both SPADE algorithms process the signal by overlapping blocks. 
Each block is processed separately and the blocks are folded back using the standard overlap--add (OLA) procedure. 
The frequency representation in each block is computed using a~twice redundant DFT (forming a~Parseval frame).
The parameters of S-SPADE and A-SPADE
are identical
and they correspond to the description in Sec.\,\ref{subsec:aspade}.
It is fairly easy and intuitive to tune them.
The algorithm works very well with default parameters ($s = 1,\ r = 1$).
During testing, we found out that for the most extreme clipping
(input SDR = 1\,dB)
it helps to increase the number of iterations slower, by incrementing $k$ every even iteration, i.e., $r = 2$.
This option lifted the average output SDR by 1.2\,dB.


The termination criterion is based on the residue being minimized,
i.e., on $A\x-\z$ for A-SPADE and on $\x - D\z$ for S-SPADE.
The algorithms run until the $\ell_2$ norm of the residue is smaller than~$\epsilon$.
We use the default $\epsilon = 0.1$ used in the original papers.
Decreasing $\epsilon$ may increase the number of iterations but does not improve the overall reconstruction quality
(in none of the considered quality measures).


\subsubsection[Plain l1 minimization]{Plain $\ell_1$ minimization}
\label{sssec:plain_l1}
The algorithms based on $\ell_1$ relaxation described in Sec.\,\ref{sec:Simple.weighted.l1.approach} are designed to process the input signal all at once using the DGT,
with the DGT parameters specified above in this section.

The synthesis variant was computed using the
DR algorithm (Alg.\,\ref{alg:DR.declipping}) and the analysis variant was solved by the
CP algorithm (Alg.\,\ref{alg:CP.for.declipping}).
It was quite difficult to tune the parameters in these algorithms,
since each test sound required a~slightly different setting to obtain reasonable convergence. 
It also happened sometimes that the output SDR started to drop after reaching its peak, and then it stabilized at a~lower value.
\edt{Nevertheless, we kept the algorithms running for the full number of iterations, since usually the values of the objective metrics still got increased even after this SDR peak point.}

In such a case, we let the algorithm reach the maximum number of iterations and we take the result from the final iteration. 
Note that the just-described behavior is
more common for methods
below that employ the same DR and CP algorithms but use coefficient weighting, i.e., $\w\not=\mathbf{1}$.
Because of these issues,
we set all the parameters to unity, which turns out to be a~compromise covering all the cases; we set $\lambda = \gamma = 1$ for the DR algorithm and $\zeta = \sigma = \rho = 1$ for the CP algorithm.

In both algorithms, the convergence criterion was set strictly to 3000 iterations, where it was certain that the algorithms reached the minimum with sufficient accuracy.


\subsubsection[Declipping in Sparselend (Rl1CC)]{Declipping in Sparseland R$\ell_1$CC}
The original codes of the R$\ell_1$CC method rely on the CVX,
whose disadvantage is that the transforms are handled
only in the form of a~matrix;
this prevents CVX from using longer window lengths.
Therefore, we re-implemented the original approach using the DR algorithm, as described in Alg.\,\ref{alg:DR.declipping.reweighted}.
\edt{We no longer used matrices but fast operator-based transforms.}
For the DR algorithm, we used the same setting as in the non-weighted case, i.e., $\lambda = \gamma = 1$,
but the maximum number of the DR iterations was set to 1000.
The original paper \cite{Weinstein2011:DeclippingSparseland} used 10 outer-cycle iterations of the R$\ell_1$CC algorithm,
but our implementation used only 6, since after the sixth iteration the performance starts to decrease rapidly.

Concerning the operators, we used the DGT instead of the DFT used for toy examples in \cite{Weinstein2011:DeclippingSparseland}.
The parameter $\epsilon$ used for updating the weights was set to $0.0001$.
Besides that, we introduced another parameter, $\delta$;
the inner cycle is terminated
if the relative change of the DGT coefficients between two subsequent iterations drops below $\delta$.
The specific value used in the tests was $\delta = 0.01$.
This modification was used just to speed up the computations and has no effect on the reconstruction quality.

On top of that, we also include the analysis variant using the CP algorithm (with $\zeta=\sigma=\rho=1$ and 1000 inner iterations).
Also in case of the analysis variant, the reconstruction quality starts to decrease  with the seventh outer iteration.


\subsubsection{Social Sparsity}

For the experiments, we used the implementation of the algorithm kindly provided by the authors of \cite{SiedenburgKowalskiDoerfler2014:Audio.declip.social.sparsity}.
For clarity, only the best performing variants are included in the evaluation, i.e., Empirical Wiener (EW) and Persistent EW (PEW).
In the case of the PEW, 
one needs to specify the size of the coefficient neighborhood in the TF plane. 
For our test case (audio at 44.1\,kHz and the DGT),
the best-performing size of the neighborhood was $3\times 7$
(i.e., 3 coefficients in the direction of frequency and 7 coefficients in time).

One needs to carefully tune the number of inner and outer iterations and the distribution of the parameter $\lambda$ during the course of iterations
(see Sec.\,\ref{sec:social.sparsity}).
In the final algorithm, we used 500 inner and 20 outer iterations with $\lambda$ logarithmically decreasing from $10^{-1}$ to $10^{-4}$.

As for the step size, the authors of \cite{SiedenburgKowalskiDoerfler2014:Audio.declip.social.sparsity}
report using $\gamma=0.9$, leading to fast convergence.
Following the codes provided by the authors,
our $\gamma$ rather develops according to the formula $\frac{k-1}{k+5}$, where $k$ is the iteration counter.
Such an approach corresponds to acceleration
in FISTA
\cite{beck2009fast}.

Sometimes it happens that the optimization gets stuck
(especially in the first couple of outer iterations) and starts to converge again with the next outer iteration (i.e., when $\lambda$ is decreased).
For this reason, we slightly modified the implementation by adding the $\delta$ threshold, used to break the outer iteration even if the maximum number of inner iterations has not been reached.
The $\ell_2$ norm of the difference between the time-domain solutions of the current and previous iteration is compared with $\delta$,
whose value was set to $0.001$.

Even though the algorithm does not produce signals consistent in any earlier defined sense,
the result of social sparsity approaches is usually not too far from the set $\Gamma$.

\subsubsection{Perceptual Compressed Sensing}
We were able to obtain neither the implementation of this method nor any example of output signals, unfortunately.
The authors say that it relies on CVX, which is not quite practical for our experiment.
Since the scores reported in
\cite{Defraene2013:Declipping.perceptual.compressed.sensing}
looked promising, we re-implemented the method using the Condat-V\~{u} algorithm, which mimics the algorithmic scheme suggested in \cite{Defraene2013:Declipping.perceptual.compressed.sensing}.
The signal is processed block by block as in the SPADE algorithms.

Our experiment includes the non-weighted variant, CSL1, the perceptually-weighted variant, PCSL1, and, inspired by very good results of quadratic weights in \cite{ZaviskaRajmicSchimmel2019:Psychoacoustics.l1.declipping}, 
an additional parabola-weighted variant, referred to as PWCSL1.

The parameters of the CV algorithm were set to
$\gamma = 0.01$, $\sigma = 1$, $\tau \approx 0.0186$, and $\rho = 0.99$
for all three mentioned variants.  
The maximum number of iterations was set to 500 for CSL1 and PCSL1 and 5000 for PWCSL1
\edt{(convergence is significantly slower in the latter case)}.

In contrast to the original article,
the \qm{official} implementation of MPEG PM~1 could not be used because it is strictly limited to 512-sample-long windows.
In this survey, our goal is to compare algorithms with the best possible settings and with the same DGT settings across all methods.
Therefore, we wanted to work with a~Hann window 8192 samples long with 75\,\% overlap and 16384 frequency channels. 
Hence, instead of the official implementation, we had to switch to a~slightly modified and simplified version of MPEG Layer~1 PM~1, which is not restricted in terms of the block length.

The PCSL1 algorithm places the original reliable samples at the reliable positions at the very end of processing.
This leads to some ODG gain for mild clipping levels.
However, as  discussed earlier, we present results without such a replacement.
%
%

\subsubsection[Psychoacoustically motivated l1 minimization]{Psychoacoustically motivated $\ell_1$ minimization}

The algorithms used here are the same as in the case of the plain $\ell_1$ minimization (Sec.\,\ref{sssec:plain_l1}), but the weights are now  perceptually motivated.
The original paper \cite{ZaviskaRajmicSchimmel2019:Psychoacoustics.l1.declipping}
presented several ways of implementing the psychoacoustical information into the declipping process
(only for the synthesis case).
The best option turned out to be simple weights that grow quadratically with frequency. 

In the experiments, we present only this \qm{parabola-weighted} option,
but we newly include the analysis variant as well.
All the parameters and the maximum number of iterations for both DR and CP are identical to the
plain case.

\subsubsection{Dictionary Learning approach}
For the dictionary learning algorithm, each signal is first decomposed into $T$ overlapping windows $\y_1, \dots, \y_T$ of size $1024$, for a total of approximately $T = 1200$ windows per signal.
These are
directly used as inputs to the algorithm, such that the dictionary is learned and evaluated on the same signal.

As the optimal sparsity parameter $K$ depends on the signal as well as on the clipping level, we adopt here an \emph{adaptive sparsity} strategy.
In the first sparse coding step \eqref{eq:consistentIHT}, we first iterate \eqref{eq:consistentIHT} with $K = 1$, then sequentially increase $K$ every few iterations, until an error threshold $\epsilon$ is reached.
The resulting sparsity level $\hat{K}$ is then fixed throughout the rest of the algorithm.
The dictionary algorithm is initialized with a~real DCT dictionary of size $P = 2048$.
We perform 20 iterations of sparse coding and dictionary update steps. The sparse coding steps (apart from the first one, which uses an adaptive sparsity strategy) are computed using 20 iterations of consistent IHT. The dictionary update step is computed using 20 steps of accelerated gradient descent.
To improve the convergence,
the sparse coefficients and dictionary are always initialized using estimates from the previous iteration.
Using the learned dictionary $\hat{D}$ and sparse coefficients $\hat{\z}_1, \dots, \hat{\z}_T$, each of the individual windows is reconstructed as $\hat{D}\hat{\z}_1, \dots, \hat{D}\hat{\z}_T$, and 
the estimated signal is then recovered using the overlap--add.

\subsubsection{NMF}
As for the NMF-based declipping, we used the STFT with the sine window of size $F = 2048$ samples with 50\% overlap.
The NMF model order was set to $K = 20$.
The GEM algorithm~\ref{alg:NMF_algorithm} was run for 50 iterations.
These choices follow those used in the corresponding paper \cite{BilenOzerovPerez2015:declipping.via.NMF} except for the STFT window size, which was 64\,ms (1024 samples for 16kHz-sampled signals) in \cite{BilenOzerovPerez2015:declipping.via.NMF}
and is 46.4\,ms (2048 samples for 44.1kHz-sampled signals) here.
Note that the computational load of the NMF approach grows drastically with increasing window length.
This is due to matrix inversion in Wiener filtering~\eqref{eq::nmf_Wiener} and to the iterative ad hoc clipping constraint management strategy described at the end of Sec.\,\ref{sec::nmf_declipping}.
As such, even with the 2048-sample-long window, declipping some sequences took more than 8 hours.
This is why we have not chosen a~longer window size for experiments.



\subsubsection{Janssen}
For the evaluation of the Janssen algorithm, we have adapted the codes published in the SMALLbox.
The signal is processed block by block.
The order of the AR filter does not play a significant role in the quality of declipping;
it turns out that the number of iterations is significantly more important. 
The \dsdr{} increases with the number of iterations,
but from a~certain point, the algorithm starts to diverge wildly.
The reason is probably that the algorithm does not have enough reliable data to fit reasonable AR parameters to the known signal.
This point of change differs among the test signals and it is highly influenced by the input SDR 
(the smaller the input SDR, the sooner the divergence appears).
In our experiments, the order of the filter is set to 512 and we run 3 iterations.
Using 5 iterations produces slightly better results, but for one of the test signals at 1\,dB input SDR the divergence appears, devaluating the average score.

\subsection{Computational cost}
\label{sec:computation.cost}
Though the survey concentrates primarily on the reconstruction quality, some remarks on the computational cost of the methods are valuable.
Below we quote the rough computational time needed to process one second of audio (i.e., 44100 samples).
No parallel computing is considered, 
and the time spent by tuning the parameters is not included.
The overall time needed to completely recompute our experiments is estimated at one month when performed on a~common
PC.

\begin{enumerate}
     
\item C-OMP: It takes 5 to 10 minutes to declip one second of the signal.

\item SPADE: The clipping threshold determines the performance---the higher the input SDR, the longer it takes for the SPADE algorithms to converge.
Processing one second of audio takes from 22 up to 64 seconds for A-SPADE and 14 up to 52 seconds for S-SPADE.

\item Plain $\ell_1$: Computational time is roughly 20 seconds
for both the DR and CP algorithms, independent of the clipping threshold.

\item Reweighted $\ell_1$: It takes 66 seconds
for the DR algorithm and 56~seconds for the CP algorithm.

\item Social Sparsity:
Slightly less than 2 minutes for EW and slightly more than 2 minutes in the case of PEW.

\item Perceptual Compressed Sensing:
CSL1 \& PCSL1: Roughly 20 seconds, PWCSL1 below three minutes.

\item PW$\ell_1$: Same as for Plain $\ell_1$.

\item Dictionary Learning:
1 to 2 minutes depending on the clipping level, the algorithm generally converging a bit faster when the clipping level is low.

\item NMF:
Average computation time is 30 minutes per one second of audio.
The particular time depends on the input SDR---for the lowest, the cost can rise up to 1~hour.

\item Janssen:
Depends heavily on the input SDR; the durations differ by two orders of magnitude:
for 1\,dB input SDR Janssen takes about 16 minutes per second of audio,
and for 20\,dB input SDR it is done in 5--15~s.

\end{enumerate}

\subsection{Results and discussion}
\label{sec:results}
Results of the declipping in terms of performance are presented and commented on in this section. 
Recall that the comparison is done in terms of \edt{four objective metrics}---\dsdrc{}, PEAQ, PEMO-Q \edt{and \Rnonlin}.
In the bar graphs that follow,
algorithms coming from the same family share the same color. 
If a~method was examined in both the analysis and the synthesis variant, the analysis variant is graphically distinguished via hatching.
Other variants (e.g., multiple shrinkage operators in the SS algorithms or different weights within the CSL family) use gray stippling.
The abbreviations used in the legends are used all over the text, but also summarized in Table~\ref{tab:alg_abbrev}.

\begin{figure*}
    \input{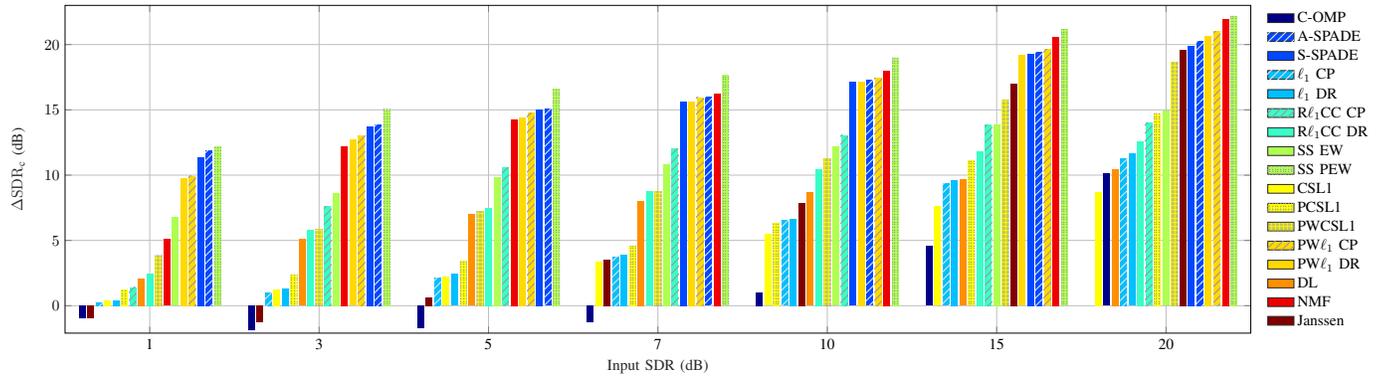}
		\vspace{-1.2em}
    \caption{%
    Average $\dsdrc$ results.
    The abbreviations from the legend are summarized in Table~\ref{tab:alg_abbrev}.
    }
    \label{fig:results_dsdrc}
\end{figure*}

\begin{figure*}
    \input{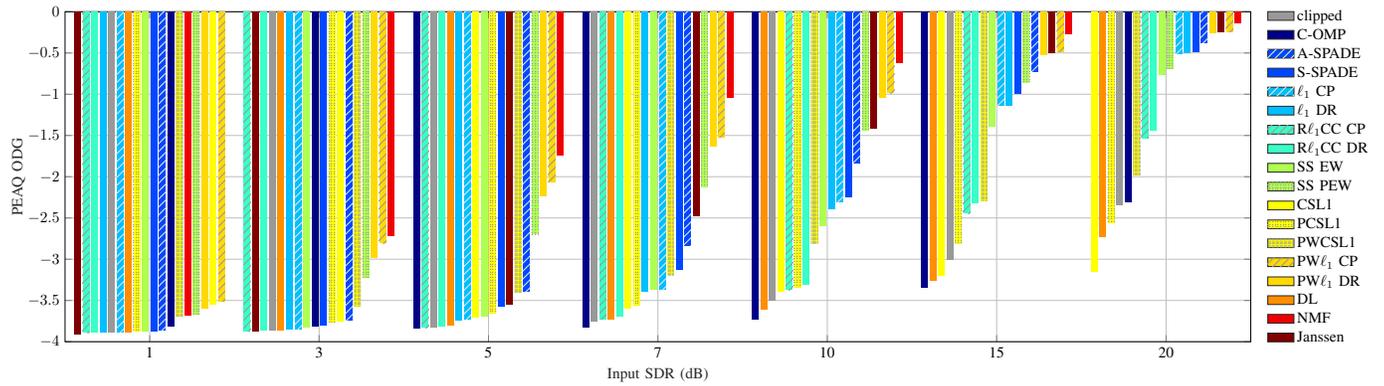}
		\vspace{-1.2em}
    \caption{Average PEAQ ODG results.
    The PEAQ ODG of the clipped signal is depicted in gray.}
    \label{fig:results_peaq}
\end{figure*}

\begin{figure*}
    \input{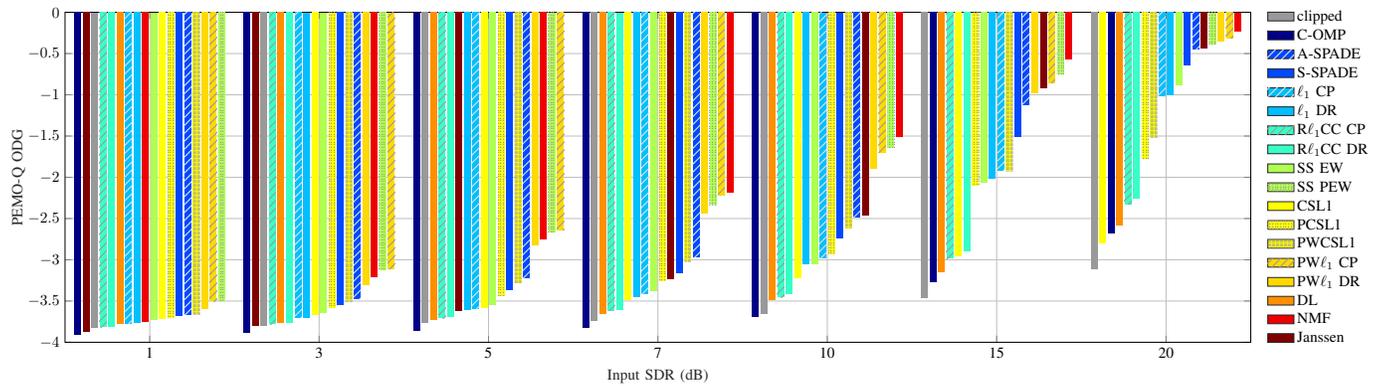}
		\vspace{-1.2em}
    \caption{Average PEMO-Q ODG Results.}
    \label{fig:results_pemoq}
\end{figure*}

The \dsdrc{} results are presented in Fig.\,\ref{fig:results_dsdrc}, PEAQ ODG values in Fig.\,\ref{fig:results_peaq}, PEMO-Q ODG values in Fig.\,\ref{fig:results_pemoq}, \edt{and \Rnonlin{} values in Fig.\,\ref{fig:results_rnonlin}}.
The reader can easily draw a~number of conclusions by studying the plots;
however, we try to summarize the most important and interesting facts inferred from these results.
We concentrate more on the ODG score designed to reflect properties of the human auditory system.




%


\begin{figure*}
    \input{figz/Results_Rnonlin_calc_sorted}
		\vspace{-1.2em}
    \caption{Average \Rnonlin{} Results.}
    \label{fig:results_rnonlin}
\end{figure*}

\begin{table*}[t]
    \footnotesize
    \centering
    \caption{Table of examined algorithms and their abbreviations and references.
    Some of the algorithms were proposed in this article in order to make it as complete as possible.
    Two algorithms are not presented.
    }
        \begin{tabular}{|Sl|Sl|Sl|Sl|}
        \hline
        \rowcolor{white!90!black}
        \multicolumn{1}{|Sc|}{Abbreviation} & 
        \multicolumn{1}{|Sc|}{Full name} & 
        \multicolumn{1}{|Sc|}{Algorithm utilized} &
        \multicolumn{1}{|Sc|}{Reference} \\ \hline\hline
        C-OMP & Constrained Orthogonal Matching Pursuit & Alg.\ \ref{alg:Constrained.OMP} & Adler'11 \cite{Adler2011:Declipping}\\ 
        \rowcolor{white!95!black}
				S-SPADE & Synthesis SParse Audio DEclipper & Alg.\ \ref{alg:sspade_dr} & Záviška'19 \cite{ZaviskaRajmicMokryPrusa2019:SSPADE_ICASSP}\\
        A-SPADE & Analysis SParse Audio DEclipper & Alg.\ \ref{alg:aspade} & Kiti\'c'15 \cite{Kitic2015:Sparsity.cosparsity.declipping}\\ 
        \rowcolor{white!95!black}
				$\ell_1$ DR & $\ell_1$ minimization using Douglas--Rachford (synthesis) & Alg.\ \ref{alg:DR.declipping} & Rajmic'19 \cite{RajmicZaviskaVeselyMokry2019:Axioms}\\
        $\ell_1$ CP & $\ell_1$ minimization using Chambolle--Pock (analysis) & Alg.\ \ref{alg:CP.for.declipping} & analysis variant of \cite{RajmicZaviskaVeselyMokry2019:Axioms} \\ 
        \rowcolor{white!95!black}
				R$\ell_1$CC DR & Reweighted $\ell_1$ min.\ with Clipping Constraints using Douglas--Rachford (synthesis) & Alg.\ \ref{alg:DR.declipping.reweighted} & Weinstein'11 \cite{Weinstein2011:DeclippingSparseland}\\
        R$\ell_1$CC CP & Reweighted $\ell_1$ min.\ with Clipping Constraints using Chambolle--Pock (analysis) & --- & analysis variant of \cite{Weinstein2011:DeclippingSparseland}\\
        \rowcolor{white!95!black}
				SS EW & Social Sparsity with Empirical Wiener & Alg.\ \ref{alg:social.sparsity} & Siedenburg'14\,\cite{SiedenburgKowalskiDoerfler2014:Audio.declip.social.sparsity}\\  
        SS PEW & Social Sparsity with Persistent Empirical Wiener & Alg.\ \ref{alg:social.sparsity} & Siedenburg'14 \cite{SiedenburgKowalskiDoerfler2014:Audio.declip.social.sparsity}\\ 
        \rowcolor{white!95!black}
				CSL1 & Compressed Sensing method minimizing $\ell_1$ norm & Alg.\ \ref{alg:CV.for.perceptual.declipping} & Defraene'13 \cite{Defraene2013:Declipping.perceptual.compressed.sensing} \\ 
        PCSL1 & Perceptual Compressed Sensing method minimizing $\ell_1$ norm & Alg.\ \ref{alg:CV.for.perceptual.declipping} & Defraene'13 \cite{Defraene2013:Declipping.perceptual.compressed.sensing} \\ 
        \rowcolor{white!95!black}
				PWCSL1 & Parabola-Weighted Compressed Sensing method minimizing $\ell_1$ norm & Alg.\ \ref{alg:CV.for.perceptual.declipping} & variant of \cite{Defraene2013:Declipping.perceptual.compressed.sensing} \\
				PW$\ell_1$ DR & Parabola-Weighted $\ell_1$ minimization using Douglas--Rachford (synthesis) & Alg.\ \ref{alg:DR.declipping} & Záviška'19b \cite{ZaviskaRajmicSchimmel2019:Psychoacoustics.l1.declipping}
        \\
        \rowcolor{white!95!black}
				PW$\ell_1$ CP & Parabola-Weighted $\ell_1$ minimization using Chambolle--Pock (analysis) & Alg.\ \ref{alg:CP.for.declipping} & analysis variant of \cite{ZaviskaRajmicSchimmel2019:Psychoacoustics.l1.declipping} \\ 
        DL & Dictionary Learning approach & Alg.\ \ref{alg:DL_algorithm} & Rencker'18 \cite{RenckerBachWangPlumbley2018:Consistent.dictionary.learning.LVA}\\
        \rowcolor{white!95!black}
				NMF & Nonnegative Matrix Factorization & Alg.\ \ref{alg:NMF_algorithm} & Bilen'15 \cite{BilenOzerovPerez2015:declipping.via.NMF} \\ 
        Janssen & Janssen method for inpainting & --- & Janssen'86 \cite{javevr86}\\ \hline
        
    \end{tabular}

    \label{tab:alg_abbrev}
\end{table*}

First of all, note that the SDR scores correlate with the ODG scores to a certain degree.
There are exceptions, however---compare, for example, R$\ell_1$CC CP and R$\ell_1$CC DR,
which behave just the opposite way (SDR versus ODG).%
\footnote{%
This corresponds to our experience with reweighting in audio inpainting,
see \cite{MokryRajmic2019:Reweighted.l1.inpainting},
but we do not have an explanation for this effect unfortunately.
}
Note also that the ODG values of PEMO-Q are uniformly worse than those of PEAQ,
but the relation between scores of the individual methods is retained.
An exception is the family of CSL1, PCSL1, PWCSL1, where we observe a~difference.

The main messages can be summarized as follows:
\begin{itemize}
    \item
    Both variants of the SPADE algorithm perform similarly and very well in terms of all the \edt{four} metrics
    and across all levels of degradation, while the analysis variant is slightly preferred.
    \item
    Using social sparsity leads to very good results.
    In particular, the SS\,PEW method (which assumes persistence of frequencies in time)
    performs overall best in terms of the SDR and is one of the best few in terms of the perceptual measures.
    \item
    In the medium to mild clipping regime, NMF is the clear winner in terms of ODG and it also performs very good in SDR.
    With more severe clipping (1 and 3\,dB) it behaves worse but is still very competitive.
    \item
    Despite its simplicity, the results of the parabola-weighted $\ell_1$ minimization are uniformly very good, in terms of all the metrics.
    Its SDR values more or less correspond to those of SPADE,
    and the ODG scores show that for wild clipping, PW$\ell_1$ is even the best declipping method.
    \item
    Introduction of reweighting improves the plain $\ell_1$ minimization, especially for the analysis variant,
    but this observation only holds for the SDR.
    In terms of ODG, the effect is reversed.
    In fact, the performance of plain $\ell_1$ can be found surprisingly satisfactory in the PEAQ ODG graph.
    \item
    The family of psychoacoustically weighted optimizations (CSL1) failed.
    The best results are achieved using the parabolic weights, which are in fact not psychoacoustically motivated.
    These observations are especially interesting since the original article
    \cite{Defraene2013:Declipping.perceptual.compressed.sensing}
    reported a~better declipping quality (but on a~different dataset).
    \item
    The low scores of dictionary learning may be probably attributed to
    the fact that it uses the IHT algorithm in its sparse coding step
    (see Alg.\,\ref{alg:DL_algorithm}), which has been recently surpassed by its successor, SPADE, for example.
    Another factor could be that the initial dictionary is the real-valued DCT and that the iterates of Alg.\,\ref{alg:DL_algorithm} remain in the real domain, causing phase-related artifacts.
    \item
    The Janssen method performs well only in the very high input SDR regime, otherwise it fails due to the lack of reliable samples.
    Recall that Janssen was included in the study since 
    it performs on par with state-of-the-art methods in audio inpainting (of compact signal gaps).
    Clearly, the hypothesis that it could be similarly successful also in declipping is not validated.
\end{itemize}

Besides the reconstruction quality, which is the main concern of the article,
other factors can also be taken into consideration.
For example, consider the rough cost of computation that has been reported in Sec.\,\ref{sec:computation.cost}.
According to these values, the NMF is 90-times slower than the PW$\ell_1$ algorithms, while producing results of almost identical quality for most clipping levels.
On the other hand, some methods require painful parameter tuning to achieve good results.
In that regard, NMF or SPADE in particular can be seen as advantageous.





\subsection{Software and data}
Besides the numerical evaluation, the intention of the article is to collect implementations of the examined methods and to make them publicly available,
both for the reproducibility purposes and to stimulate future research in this area.
The webpage 
\begin{center}
	https://rajmic.github.io/declipping2020/
\end{center}
contains the link to the MATLAB implementation
(except the NMF,
which is not publicly available).
%
The tests have been performed in MATLAB version 2019b.

This article provides an objective evaluation through PEAQ and PEMO-Q, which pursue being as close as possible to human perception.
Individual subjective assessment, which is always the most decisive,
can be made via the supplied webpage, where all 
sound examples are directly playable (or downloadable).

\section{Conclusion and Perspectives}
The article presented the declipping problem and an overview of methods used to solve it.
Besides such a~survey, several popular declipping methods of different kinds were selected for deeper evaluation.
This is the first time so many methods are compared based on the same audio dataset
(moreover sampled at 44.1\,kHz).
The main focus of the article is on the reconstruction quality, which is reported in terms of \edt{four} metrics.
However, other factors like the computation cost and complexity in tuning parameters are also discussed.

The algorithms studied and compared in this paper exhibit various performances and computational times.
Some algorithms perform better at low clipping levels, while others perform better at high clipping levels.
The choice of an algorithm thus depends on the input data.
Nevertheless, the methods based on social shrinkage, nonnegative matrix factorization, weighting the transform coefficients and, last but not least, the SPADE seem to yield results that make them preferred choices;
\edt{in most conditions, they scored significantly better than the majority of other methods in terms of the perceptual metrics.}
Depending on the application, the computational time of each algorithm and the time-consuming tuning of parameters might also be a~decisive selection criterion.
\edt{From these viewpoints, variants of the SPADE are attractive.}

Directions for future research may include combining strategies and modeling assumptions of the various algorithms presented in this paper.
For instance, the social sparsity regularizer of \cite{SiedenburgKowalskiDoerfler2014:Audio.declip.social.sparsity} or the perceptual weights of \cite{Defraene2013:Declipping.perceptual.compressed.sensing} could be combined with S-SPADE or other algorithms.
Dictionary learning could be combined with S-SPADE or social sparsity.
Most algorithms discussed here use the synthesis model, whiledeveloping their \emph{analysis}\/ counterpart could also be a~promising idea.
We could also imagine assigning weights, in order to favor clipping consistency.
Finally, we have focused in this paper on unsupervised techniques, which do not rely on a training set with clean data.
However, the success of supervised techniques, and in particular deep learning based techniques, in tackling many other problems in computer vision, speech recognition, and audio analysis, motivates further study of supervised techniques in audio declipping.
Recent deep learning approaches to audio declipping have shown promising results in the context of speech declipping \cite{Bie2015:Detection.and.Reconstruction.of.speech.for.speaker.recongnition, kashani2019:Image2Image.CNN.Speech.Declipping,MackHabets2019:Declipping.Speech.Using.Deep.Filtering}.
A potential research direction would be to combine the power of supervised techniques with signal assumptions, modeling and algorithms discussed in this article.
%
One of such directions could be the recent finding that artificial networks
that bear the structure of unfolded proximal algorithms are able to join the signal modeling and learning from data,
possibly keeping advantages of both distinct worlds,
see for example 
\cite{Chang2017:One.Network.to.Solve.Them.All}
in the context of image processing.

Note that this survey and the papers under consideration only investigate declipping of signals that are clipped in the digital domain.
However, when clipping occurs in the analog domain,
it is different since before the A/D conversion,
a~low-pass filter is applied to avoid aliasing.
Since the clipping distortion is wide-band, clipping aliasing \cite{kraght2000aliasing}, quite an unpleasant distortion, is present in the latter case.
This effect is well-known to audio engineers, and a~digital aliasing-free clipping or compression of dynamics is often performed via upsampling \cite{mapes1998worst}.
For example, \cite{esqueda2016aliasing} addresses the aliasing reduction in digitally-clipped signals, though without declipping itself.
While the methods covered by this survey reduce both aliasing and the remaining narrow-band clipping distortion,
if the clipping is carried out in the analog domain or in a particular aliasing-free manner (e.g., via upsampling \cite{mapes1998worst}),
different new declipping algorithms should be developed and applied, simply because the clipping process is different in that case.
\section*{Acknowledgment}
The authors would like to thank M.\,Kowalski for the implementations related to the paper \cite{SiedenburgKowalskiDoerfler2014:Audio.declip.social.sparsity},
to O.\,Mokrý for computing the results of the Janssen method,
to Z.\,Průša for helping with the accompanying HTML page.
Thanks to S.\,Kitić and N.\,Bertin for discussing SPADE algorithms and projections with tight frames.
\edt{Thanks also to F.\,Ávila for the discussion on the \Rnonlin{} metric.}

The work of P.\,Záviška and P.\,Rajmic was supported by the Czech Science Foundation (GA\v{C}R) project number 20-29009S. The work of L.\,Rencker was supported by the European Union’s H2020 Framework Programme (H2020-MSCA-ITN-2014) under grant agreement no.\ 642685 MacSeNet.


\ifCLASSOPTIONcaptionsoff
  \newpage
\fi



%
%
%

\newcommand{\noopsort}[1]{} \newcommand{\printfirst}[2]{#1}
  \newcommand{\singleletter}[1]{#1} \newcommand{\switchargs}[2]{#2#1}

\end{document}